\documentclass[aps, twocolumn, showkeys, longbibliography]{revtex4-1}

\usepackage{graphicx}
\usepackage{dcolumn}
\usepackage{bm}

\newcommand{\dograph}[2]{
	\begin{figure}
	\includegraphics[width=3in]{#1}
	\caption{\label{fig:#1}#2}
	\end{figure}
}

\begin{document}

\title{Phase Coherence and Andreev Reflection in Topological Insulator Devices}
\author{A.D.K.~Finck$^{1}$, C.~Kurter$^{1}$, Y.S.~Hor$^2$, D.J.~Van Harlingen$^1$}
\affiliation{$^1$Department of Physics and Materials Research Laboratory, University of Illinois at Urbana-Champaign, Urbana, Illinois 61801
\\
$^2$Department of Physics, Missouri University of Science and Technology, Rolla, MO 65409}
\date{\today}

\begin{abstract}

Topological insulators (TIs) have attracted immense interest because they host helical surface states.  Protected by time-reversal symmetry, they are robust to non-magnetic disorder.  When superconductivity is induced in these helical states, they are predicted to emulate $p$-wave pairing symmetry, with Majorana states bound to vortices.  Majorana bound states possess non-Abelian exchange statistics which can be probed through interferometry.  Here, we take a significant step towards Majorana interferometry by observing pronounced Fabry-Perot oscillations in a TI sandwiched between a superconducting and normal lead.  For energies below the superconducting gap, we observe a doubling in the frequency of the oscillations, arising from an additional phase from Andreev reflection.  When a magnetic field is applied perpendicular to the TI surface, a number of very sharp and gate-tunable conductance peaks appear at or near zero energy, which has consequences for interpreting spectroscopic probes of Majorana fermions.  Our results demonstrate that TIs are a promising platform for exploring phase-coherent transport in a solid-state system.

\end{abstract}
\pacs{PACS?} 

\maketitle

\section{Introduction}

When electrons travel far distances without scattering off of impurities, their wave-like nature becomes apparent through signatures of interference.  A prominent example is ballistic transmission through a barrier between two semi-infinite leads.  Imperfect transmission through the interfaces causes partial reflection of impinging electrons.  Within the barrier, constructive interference of reflected waves leads to periodic modulation of total transmission through the barrier, resulting in resonant transmission whenever the electron wavelength is an integer multiple of twice the barrier length.  This is equivalent to $k_F L = \pi n$, where $k_F$ is the Fermi wave vector, $L$ is the barrier length, and $n$ is a nonzero integer.

This quantum phenomenon, known as Fabry-Perot interference due to its similarity to the eponymous classical optical effect, has been observed in a variety of mesoscopic systems, including nanotubes \cite{Nature.411.665}, nanowires \cite{NanoLett.10.3439}, and graphene \cite{Miao14092007}.  In topological systems with protected edge states \cite{RevModPhys.82.3045}, interferometry takes on a new significance as particles traveling through the edge states can acquire an additional phase if their paths enclose exotic quasiparticles known as anyons \cite{RevModPhys.80.1083}.  A class of anyons known as Majorana bound states (MBSs) can be realized in a TI with induced superconductivity \cite{PhysRevLett.100.096407, Alicea2012, Beenakker2013}. MBSs can encode quantum information non-locally, realizing topologically protected qubits whose states can be read out through interferometry \cite{PhysRevLett.102.216403, PhysRevLett.102.216404, PhysRevLett.103.237001}.  Recently, there has been much experimental progress in realizing superconductivity in topological systems through the study of Josephson junctions constructed on the surface of various 3D TIs \cite{PhysRevB.84.165120, NatCommun.2.575, SciRep.2.339, NatMat.11.417, NatCommun.4.1689, PhysRevX.3.021007, Orlyanchik2013, Kurter2013, PhysRevB.89.134512, PhysRevB.90.014501}.

In this paper, we describe direct studies of phase coherence and Andreev reflection in TI devices.  In each our samples, a thin flake of the 3D TI Bi$_2$Se$_3$ is placed in contact with both a superconducting lead and a normal metal lead.  The contacts are highly transparent, permitting Andreev reflection between the Bi$_2$Se$_3$ and the superconducting lead.  The chemical potential of the Bi$_2$Se$_3$ is tuned with electrostatic gating.  In more resistive devices, we find evidence of Fabry-Perot oscillations, demonstrating gate-tunable, phase-coherent transport in a TI.  For source-drain biases below the energy gap of the superconducting lead, we observe a doubling of the frequency of the conductance oscillations.  This provides firm evidence of de Gennes-Saint-James resonances \cite{PhysLett.4.151, PhysRevLett.16.453}, in which an additional phase accumulated from Andreev-reflected holes changes the period of geometric resonances by a factor of two.  When a large magnetic field is applied perpendicular to the TI surface, an intricate set of low-energy resonances emerges. These features include zero energy peaks that can persist for a range of chemical potential and magnetic field, sharing many properties of anomalies observed in nanowires \cite{Mourik25052012, Deng2012, NPhysics2479, PhysRevLett.110.126406, PhysRevB.87.241401}.  Although such anomalies were interpreted in terms of MBSs, the coexistence of the low energy resonances with phase-coherent signatures at high magnetic fields in our devices suggest alternative explanations are possible, including weak anti-localization \cite{NewJournalPhysics.14.125011}.  Our results encourage further studies of gate-tunable phase coherence in Bi$_2$Se$_3$ in order perform interferometric searches for MBSs. For example, it is necessary to discern what role do topological surface states play in the Fabry-Perot oscillations.

\section{Experimental Methods}

Single crystals of Bi$_2$Se$_3$ were grown by melting a mixture of pure Bi and Se in a stoichiometric ratio of 1.9975:3 (Bi:Se) in a vacuum quartz tube at 800 $^{\circ}$C.  Thin flakes (7-20 nm) of Bi$_2$Se$_3$ were exfoliated onto silicon substrates covered by a 300 nm thick SiO$_2$ layer.  Such thin flakes typically have a 2D carrier density of $N_{2D} \approx10^{13} - 10^{14}$ cm$^{-2}$ and low temperature mobility $\mu \approx 10^2 - 10^3$ cm$^{2}$/V-s, as determined from measurements of separate Hall bar devices with similar thicknesses as the devices mentioned in the main section and Supplemental Material \footnote{See Supplemental Material for data from additional samples and additional micrographs of the main sample}.  For estimates of the Fermi velocity, we use the value $v_F = 4.2 \times 10^5$ m/s from ARPES measurements of similar crystals of Bi$_2$Se$_3$ \cite{Nat.Phys.6.960}.  Weak anti-localization measurements of the Hall bar devices give typical phase-coherence lengths of $\ell_{\phi}= 300$ - 1000 nm at 10 mK.  For the Andreev reflection devices, superconducting leads were defined by conventional e-beam lithography and a subsequent DC sputtering of 50 nm of Nb at room temperature.  Tunnel junctions of Au/Al$_2$O$_3$/Nb reveal a superconducting gap of $\Delta = 1.5$ meV for our Nb films immediately after sputtering; in top gated TI-Nb devices, the inverse proximity effect and additional nanofabrication processing will likely reduce the gap below this pristine value.  Normal metal leads were deposited through e-beam evaporation of 5 nm of Ti and 50 nm of Au.  Brief Ar ion milling is employed before metallization \emph{in situ} to ensure good contact between the Bi$_2$Se$_3$ and the leads.  This process yields reliably transparent contacts, as demonstrated by the consistent observation of supercurrents in topological insulator Josephson junctions \cite{NatCommun.4.1689, Orlyanchik2013, Kurter2013, PhysRevB.90.014501}.  Contact resistance is estimated to be much less than 100 ohms and is thus negligible.  Typical Nb-Au lead separation is 100-250 nm.  Applying a bias to the silicon substrate permits back gating.  A top gate is created by covering the sample with 30 nm of alumina via ALD and deposition of Ti/Au over the exposed Bi$_2$Se$_3$.  The device discussed in the main section was covered with alumina but lacked a top gate.  The devices were thermally anchored to the mixing chamber of a cryogen-free dilution refrigerator equipped with a vector magnet and filtered wiring.  We perform low frequency transport measurements with standard lockin techniques, typically with a 10 nA AC excitation at $f=73$ Hz.  Unless stated otherwise, all measurements were performed at a base mixing chamber temperature of 20 mK.

\section{Transport in Zero Field}

\subsection{Andreev Reflection and Re-entrant Resistance Effect}

We study Andreev reflection in thin ($\sim 10$ nm) flakes of the 3D topological insulator Bi$_2$Se$_3$ connected to both superconducting and normal metal leads separated by $L = 100-250$ nm.  An example of a device with $L = 230$ nm is pictured in Fig.~\ref{fig:overview1}a.  Either top or back gates are used to tune the chemical potential of the surface states, depicted in Fig.~\ref{fig:overview1}b.  Here we focus on one sample, with qualitatively similar results obtained from others \cite{Note1}.  At low temperature, we observe an enhancement in conductance at source-drain biases ($V$) below the niobium gap, as expected for Andreev reflection (Fig.~\ref{fig:overview2}a) \cite{PhysRevB.25.4515, PhysRevB.85.104508, PhysRevB.86.144508}.  Near $V=0$, however, there is a downturn in the conductance.  This is consistent with the re-entrant resistance effect \cite{Artemenko1979771, PhysRevLett.76.823}, in which Andreev reflection is cancelled out by the proximity effect in the limit of zero energy and zero temperature \cite{PhysRevLett.76.823, PhysRevB.55.1123}.  Maximum conductance occurs at either a characteristic temperature or $V$ that is governed by the Thouless energy $E_T = \hbar D / L^2$, where $D$ is the diffusion constant and $L$ is the separation between the superconducting and normal metal leads on the bismuth selenide.  Magnetotransport and ARPES measurements \cite{Nat.Phys.6.960} allow us to calculate $E_T$ through $D = \frac{1}{2} v_F l = \frac{1}{2} v_F \mu (\frac{h}{2e})\sqrt{N_{2D}/\pi} \approx  0.0025$ m$^2/$s, where $v_F$ is the Fermi velocity, $l$ is the mean free path, $\mu$ is the mobility, $N_{2D}$ is the 2D carrier density.  This results in $E_T$ = 30 $\mu$V for $L=230$ nm, consistent with our transport measurements ($T_{max} = 900$ mK and $V_{max} = 85$ $\mu$V).  At high temperature (Fig.~\ref{fig:overview2}b), the conductance dip disappears and only a broad peak from Andreev reflection remains.  We note in passing a remarkable similarity between Fig.~\ref{fig:overview2}a and the predictions of Ref.~\cite{PhysRevLett.108.107005}, which posited that anomalous Andreev bound states on the surface of a topological superconductor would generate an asymmetric zero bias tunneling conductance peak \cite{PhysRevLett.107.217001} that splits at low temperature. However, we feel that our data is most likely a result of the re-entrant resistance effect.  Devices with smaller $L$ possess larger $E_T$ and are observed to have a broader suppression in conductance around zero bias \cite{Note1}.

\dograph{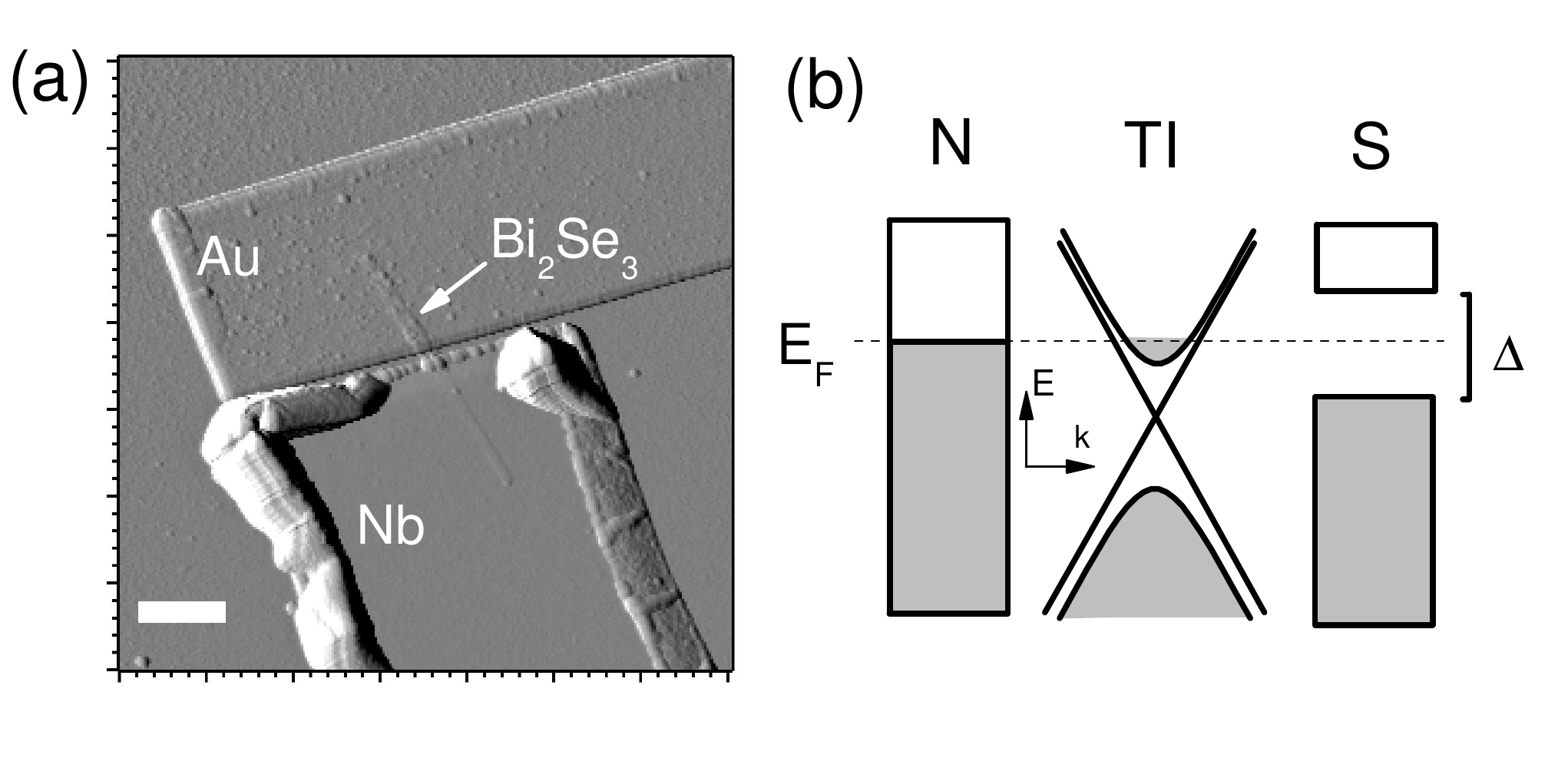}{(color online) (a) AFM image of N-TI-S device.  The Bi$_2$Se$_3$ segment originally consisted of two adjacent nanoribbons that are each 9 nm thick, but subsequent AFM imaging after measurements revealed only one nanoribbon remaining under the niobium lead \cite{Note1}.  Separation between N and S leads is $L = 230$ nm, as determined from SEM. White scale bar is 1 micron.  (b) Energy diagram for the three materials in the device.  The TI (whose dispersion relation is depicted with both gapless helical surface states and gapped trivial states) is sandwiched between a compressible normal metal and a superconductor with an energy gap $\Delta$.}

\dograph{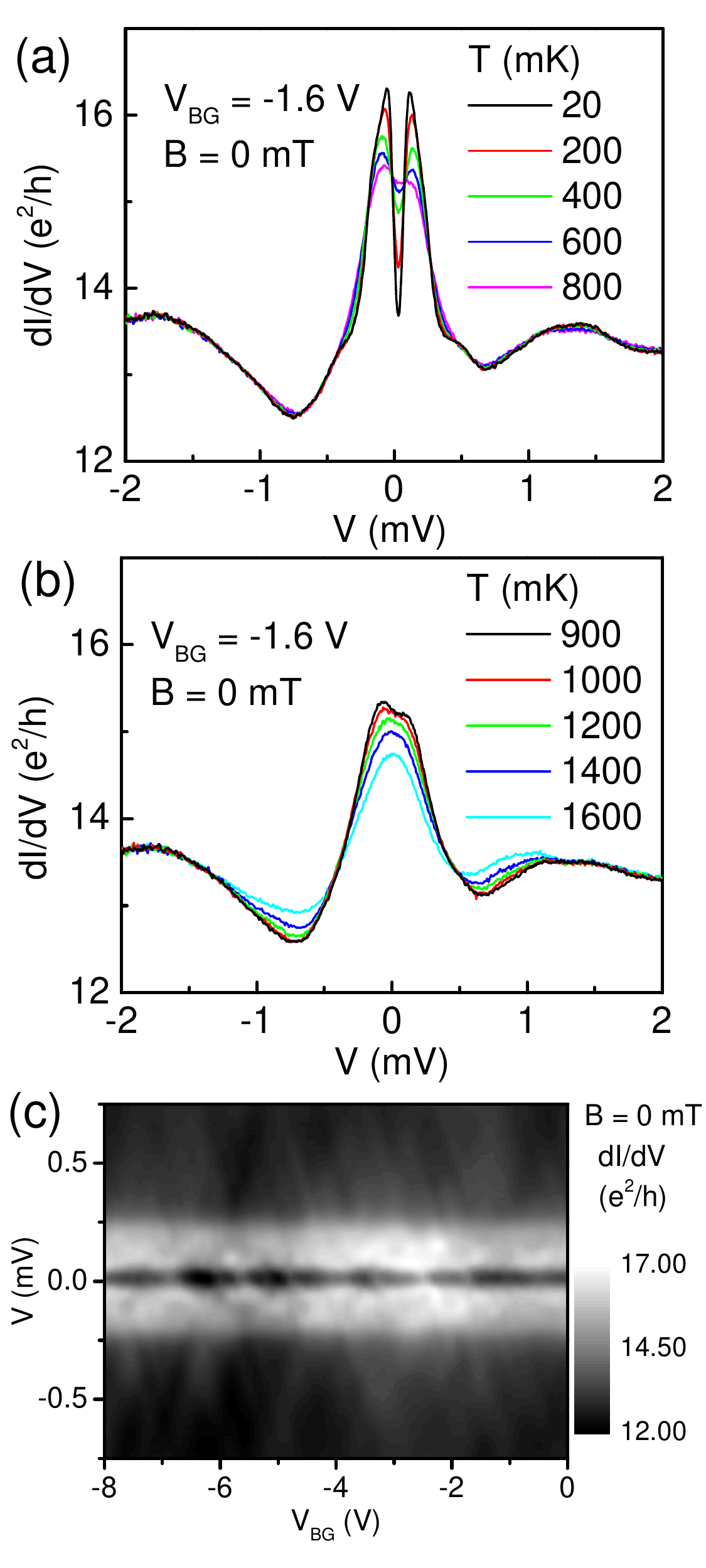}{(color online) (a) Conductance vs source-drain bias $V$ for various temperatures at zero field.  At low bias, the conductance is suppressed due to the re-entrant resistance effect. (b) Conductance traces at zero field and elevated temperature.  When the temperature exceeds the Thouless energy, the low bias conductance suppression is replaced by a broad conductance peak. (c) Color plot of $dI/dV$ vs source-drain bias $V$ and back gate bias $V_{BG}$ at zero field and 20 mK.}


\subsection{Fabry-Perot Oscillations}

To demonstrate phase-coherent transport, we turn to gate-tuned measurements.  In Fig.~\ref{fig:overview2}c, we display a color plot of conductance vs $V$ and back gate voltage $V_{BG}$.  At higher bias (Fig.~\ref{fig:fabryperotfig}a), we observe a checkerboard pattern that is characteristic of Fabry-Perot oscillations when there is a finite probability of backscattering at the interfaces between the topological insulator segment and the metallic leads.  Such backscattering could either result to a finite barrier or Fermi velocity mismatch, as well as the mixing of spin states in order to satisfy the spin-momentum locking from strong spin-orbit coupling.  The periodic variation in conductance coexist with aperiodic universal conductance fluctuations (UCFs), which have been previously reported in Bi$_2$Se$_3$ \cite{PhysRevLett.106.196801}.  Whenever the Fermi wave vector $k_F$ is tuned to a Fabry-Perot resonance, constructive interference generates enhanced conductance.  Shifts in $V$ or $V_{BG}$ will shift the Fermi energy $E_F$ of incident electrons by a proportional amount.  With a linear approximation $E_F = \hbar v_F k_F$, one expects subsequent Fabry-Perot resonances to differ in energy by $\Delta E = \frac{h v_F}{2 L}$.  For $V > \Delta / e$ (Fig.~\ref{fig:fabryperotfig}b), we observe oscillations with period $\Delta E = 0.8$ mV.  This somewhat disagrees with the estimate 3.8 mV, possibly due to renormalization of the Fermi velocity from electron-electron interactions or strong coupling to a superconductor.  We note that while previous studies have provided evidence of phase coherence in topological insulators through weak anti-localization (e.g. Ref. \cite{NatComm.4.2040}), here we demonstrate direct, gate-tunable signatures of phase-coherent transport.  Devices with smaller $L$ are observed to have larger $\Delta E$ \cite{Note1}. Remarkably, such oscillations can be observed without special care for cleanliness, as achieved in other materials through current annealing or isolating the material from dielectric impurities by suspension from the substrate \cite{Bolotin2008351, NanoLett.10.3439}.  This is likely due to the topological protection of the helical surface states \cite{RevModPhys.82.3045}, although we do not rule out the contribution from trivial states \cite{NatComm.1.128}.  The peak-to-peak amplitude of the oscillations are $\approx 0.2 e^2$/h, independent of normal state resistance, which varies from 600 $\Omega$ to 21 k$\Omega$ in our three devices.  This suggests that while bulk carriers might contribute to the zero bias conductance, the quantum interference is generated by a fixed number of 2D channels from the surface states, whose transport contributions are expected to be largely independent of bulk conditions \cite{PhysRevLett.109.116804}.

\dograph{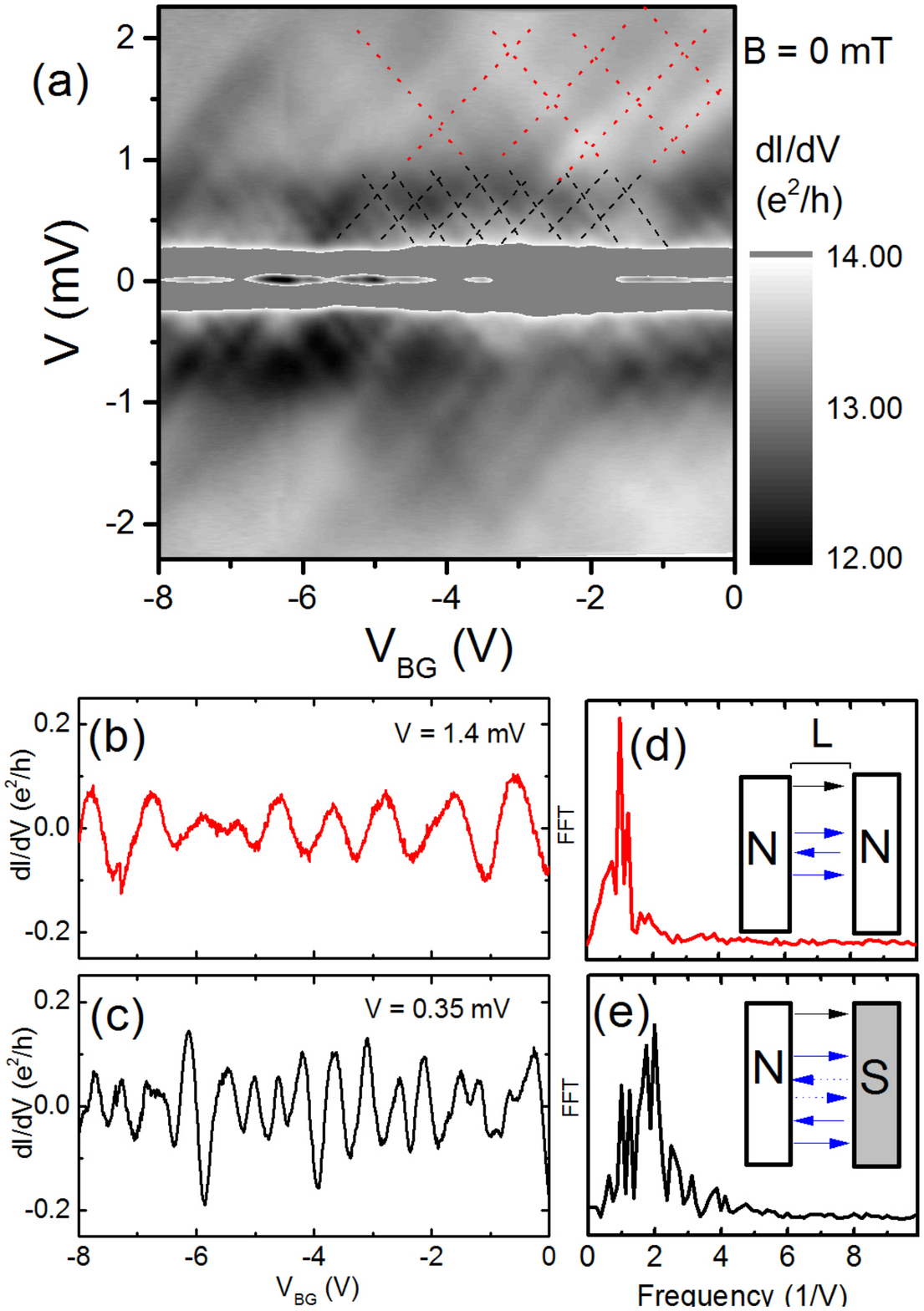}{(color online) (a) Color plot of $dI/dV$ vs source-drain bias $V$ and back gate bias $V_{BG}$ accentuating the high bias features.  Red dotted lines are guides to the eye to indicate the checkerboard pattern for conventional Fabry-Perot oscillation at high bias.  Black dashed lines follow the low bias features for de Gennes-Saint-James resonances.  (b)-(c) Plots of differential conductance vs back gate bias for two different values of source-drain bias $V$; in both cases we have subtracted a slowly-varying background.  Panel (b) shows the Fabry-Perot oscillations at high bias, with period $\Delta V_{BG} \approx 1$ V; corresponding FFT is shown in panel (d).  Panel (c) shows the low energy de Gennes-Saint-James resonances with period $\Delta V_{BG} \approx 0.5$ V; corresponding FFT is shown in panel (e).  The cartoons in insets of (d) and (e) symbolize the two sets of particle trajectories for Fabry-Perot resonances (with electrons only) and de Gennes-Saint-James resonances (mixture of holes and electrons), respectively.  Solid arrows represent the trajectories of electrons while dashed arrows represent those of holes.  Transmitted particles are shown in black while reflected particles are shown in blue.}

\subsection{Interplay between Geometric Resonances and Andreev Reflection}

The most unusual feature of the Fabry-Perot oscillations in Fig.~\ref{fig:fabryperotfig} is a doubling of their frequency when $V$ is less than the superconducting gap of niobium, as shown in Fig.~\ref{fig:fabryperotfig}c.  Fourier transforms of the oscillations (Figs.~\ref{fig:fabryperotfig}d and e) confirm this frequency doubling via a shift in spectral weight from $\Delta V_{BG} = 1$ V to $\Delta V_{BG} = 0.5$ V.  Coulomb repulsion can split otherwise degenerate levels \cite{NatMater.4.745}, but the high degree of coupling between the leads and the TI should minimize charging effects.  Furthermore, the peaks in conductance at low bias are evenly separated in $V_{BG}$, unlike the uneven spacing expected for split degeneracies.  Instead, this behavior can be understood as a modification of phase-coherent transport by Andreev reflection. Low energy electrons passing from the cavity to the superconducting lead will be retro-reflected as a hole with high probability in the absence of significant barrier, depicted in the cartoon in the inset of Fig.~\ref{fig:fabryperotfig}e.  As first pointed out by de Gennes and Saint-James, the reflected hole traverses the cavity and reflects off the interface with the normal metal lead.  The holes do not interfere with the incident electrons; instead, the hole returns to the superconducting lead to undergo Andreev reflection again, leading to an emitted electron.  This reflected electron can traverse the cavity and subsequently interfere with incident electrons.  Because in this process the reflected particles travel four times the cavity length rather than twice the cavity length, the corresponding geometric resonances occur with twice the frequency.  Although this interplay between Andreev reflection and geometric resonances was predicted over 50 years ago \cite{PhysLett.4.151, PhysRevLett.16.453}, until recently there has only been indirect evidence of this effect in the form of low-energy conductance peaks \cite{PhysRevLett.87.216808, ApplSurfSci.144.575} without any chemical potential tuning or associated observation of high-energy resonances corresponding to Fabry-Perot oscillations, which would identify them as geometric resonances.  Here, we observe a clear cross-over from high frequency geometric resonances below the energy gap to low frequency resonances above the gap, thus providing unambiguous evidence of de Gennes-Saint-James resonances.

\dograph{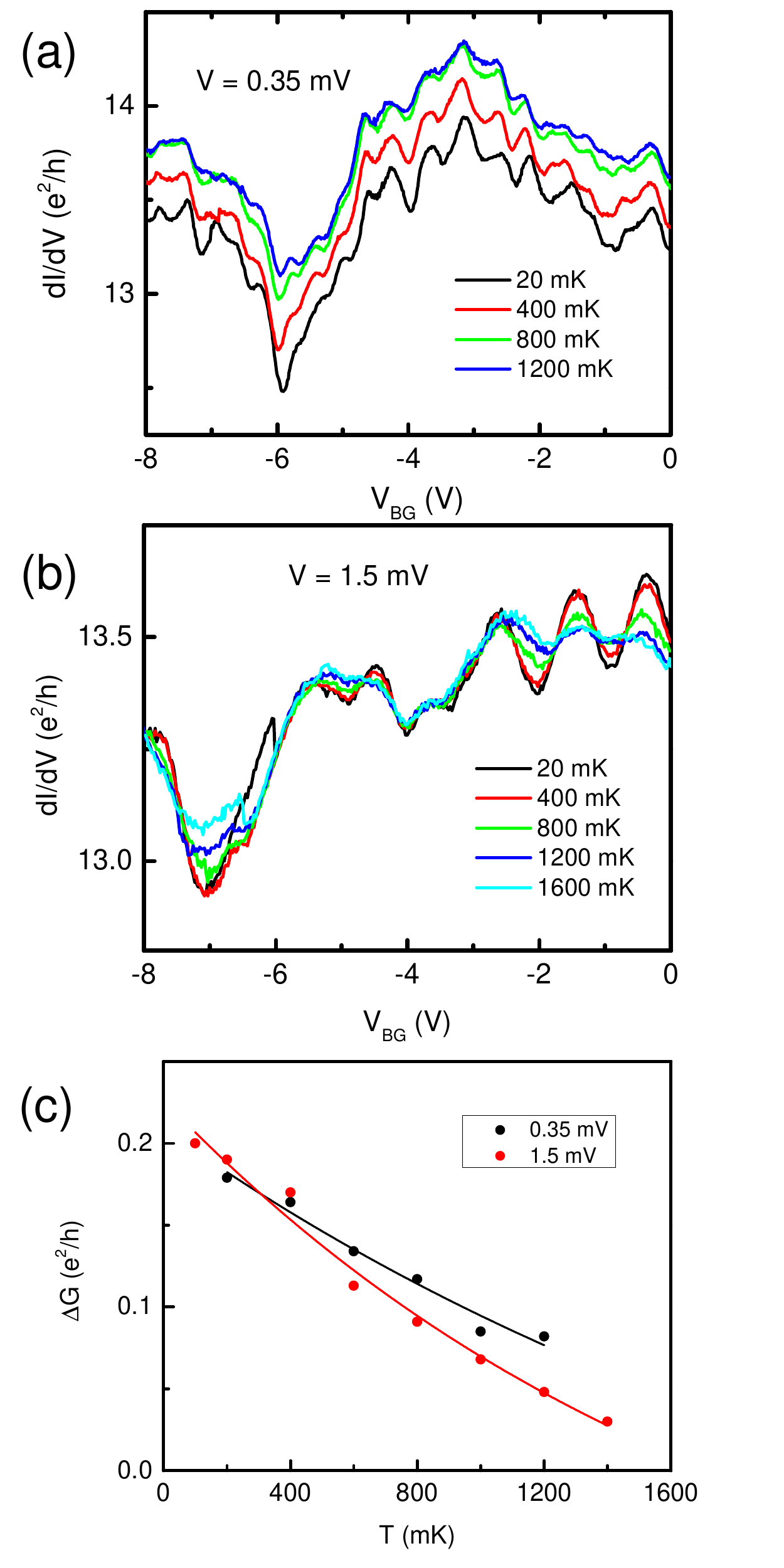}{(color online) Temperature dependence of geometric resonances for source-drain bias (a) $V = 0.35$ mV and (b) $V=1.5$ mV. (c) Peak-to-peak amplitude $\Delta G$ of geometric resonances versus temperature (circles) for two different source-drain biases, accompanied by exponential fits (solid lines).}

In Fig.~\ref{fig:fabryperottemp}a, we explore the temperature dependence of the de Gennes-Saint-James oscillations.  The amplitude of the oscillations is rapidly suppressed by thermal fluctuations, decreasing by roughly a factor of two between 20 mK and 1.6 K.  A similar suppression is observed for the Fabry-Perot oscillations at high source-drain bias (Fig.~\ref{fig:fabryperottemp}b).  We see in Fig.~\ref{fig:fabryperottemp}c that the amplitude of the oscillations show an exponential decay with temperature (i.e. $\Delta G \approx e^{- a T}$), consistent with a dephasing length that scales with $T^{-1}$ (assuming $\Delta G \approx e^{-b L / L_{\phi}}$).  This behavior has been previously seen in other open and ballistic mesoscopic systems \cite{PhysRevB.51.18037, PhysRevLett.81.200, PhysRevB.64.045327, PhysRevB.81.035312}, helping to confirm that the observed oscillations are an interference phenomenon.  We note that while weak anti-localization measurements in separate Hall bar devices show qualitatively similar sensitivity to temperature \cite{Note1}, these larger devices exhibit $L_{\phi} \approx T^{-0.5}$ \cite{SciRep.2.726}.

Our observation of Fabry-Perot oscillations also helps us to understand the behavior of planar Josephson junctions fabricated on the surface of TIs.  We have constructed Josephson junctions with similar lead separation on these Bi$_2$Se$_3$ flakes \cite{Kurter2013, PhysRevB.90.014501}, whose critical currents also vanish between 1.6 K and 2 K.  The critical current for these TI Josephson junctions as well as those from other groups \cite{NatMat.11.417, PhysRevB.89.134512} were found to rapidly increase with decreasing temperature, which was interpreted to be a sign of ballistic behavior.  Our more direct evidence of ballistic transport through Fabry-Perot oscillations helps to confirm this interpretation.

We emphasize that none of the transport signatures presented so far are unique to topological systems.  Crystals of Bi$_2$Se$_3$ are known to possess both topological surface states as well as trivial states in the bulk from unintentional doping \cite{RevModPhys.82.3045} and at the surface due to band-bending \cite{NatComm.1.128}.  The trivial states can contribute to electrical conductance and potentially exhibit phase coherence like in a conventional semiconductor.  Furthermore, there remains the question of the role of the top and bottom topological surface states.  We observe only one set of geometric resonances in our devices, independent of whether a bias is applied to the back gate or a top gate \cite{Note1}.  We do not observe any gate-independent resonances that remain at fixed source-drain bias.  This is consistent with phase-coherent transport through both upper and lower surface states whose chemical potentials are locked together due to large interlayer electrostatic coupling \cite{NatPhys.8.460}.

\section{Phase-coherent Transport in a Magnetic Field}

We next consider transport in magnetic fields applied perpendicular to the TI surface.  While in-plane fields have no influence up to at least 100 mT, with out-of-plane fields we observe a steady evolution of the background signal from universal conductance fluctuations and suppression of superconductivity, as shown in Fig.~\ref{fig:fabryperotb}a and b.  While the amplitude of the Fabry-Perot oscillations are similarly reduced, their phase does not change appreciably, suggesting that the fields we use are too small to observe the phase shift expected from Klein backscattering \cite{PhysRevLett.101.156804}.  Small shifts can be observed in the de Gennes-Saint-James resonances (Fig.~\ref{fig:fabryperotb}a), but it is difficult to reliably disentangle them from the evolving background signal due to universal conductance fluctuations.

\dograph{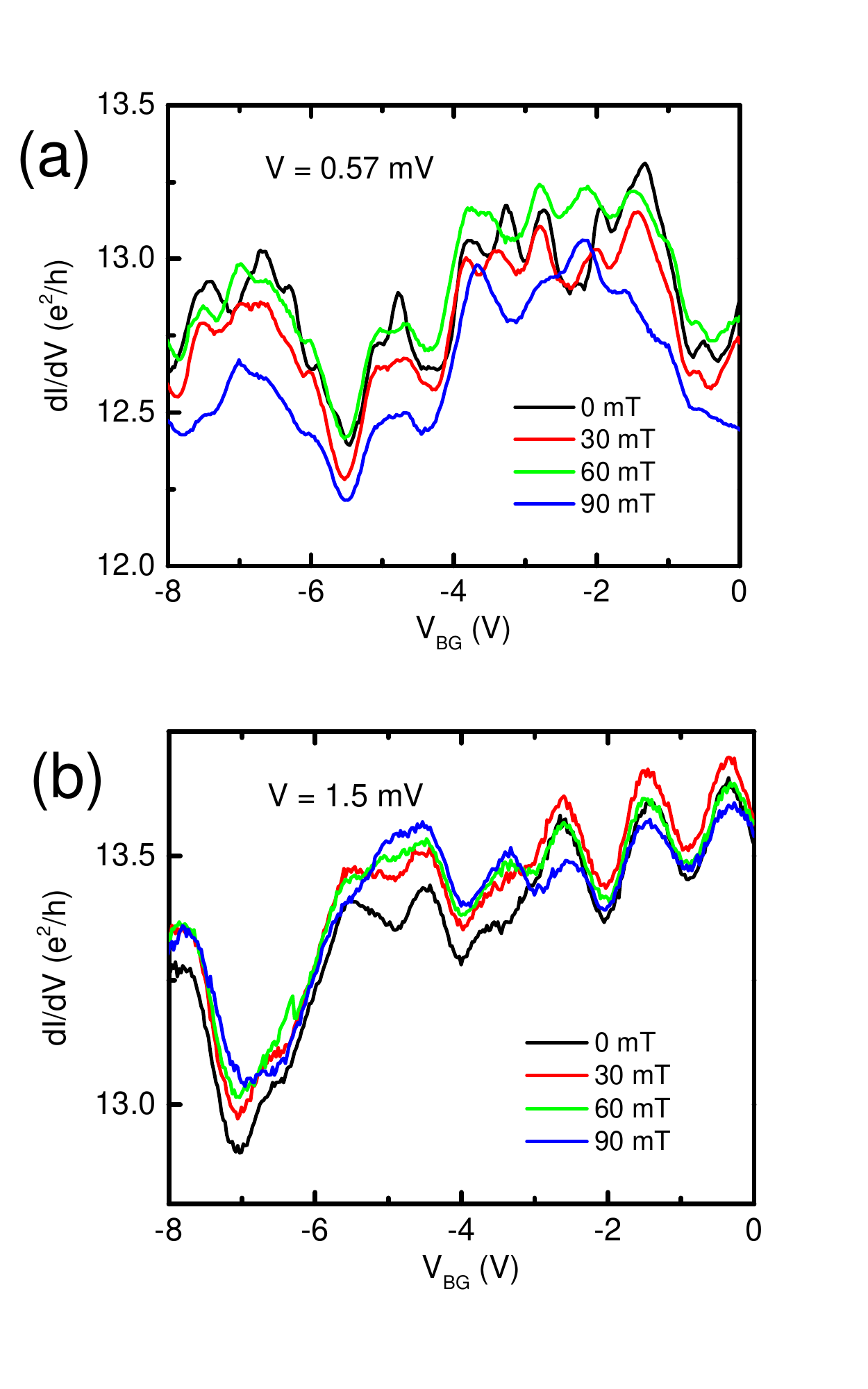}{(color online) Magnetic field dependence of geometric resonances for source-drain bias (a) $V = 0.35$ mV and (b) $V=1.5$ mV.}

\subsection{Emergence of Low-Energy Resonances at High Field}

Strikingly, in two of our samples we observe a number of narrow, low energy conductance peaks emerge at large magnetic fields.  An example is shown in Fig.~\ref{fig:transport80mT}a, with data from the second device shown in the Supplemental Material \cite{Note1}.  The location of the peaks can be tuned readily with $V_{BG}$, causing them to even form a zero bias conductance peak for certain gate ranges with height $\approx 0.2 e^2/h$, indicated in Fig.~\ref{fig:transport80mT}b.  The peaks split and reform in a quasi-periodic fashion that does not seem to be directly related to the Fabry-Perot oscillation period. They are also significantly narrower than even the de Gennes-Saint-James resonances, with source-drain widths of $\Delta V \approx 50-60$ $\mu$V according to FWHM.  In Fig.~\ref{fig:zba}a, we detail the field evolution of one of these zero bias anomalies.  We find that the zero energy resonance forms at finite field and can persist for at least up to 120 mT.  At elevated temperatures (Fig.~\ref{fig:zba}b), we find these anomalies broaden and merge with the background due to a combination of thermal fluctuations and dephasing.

\dograph{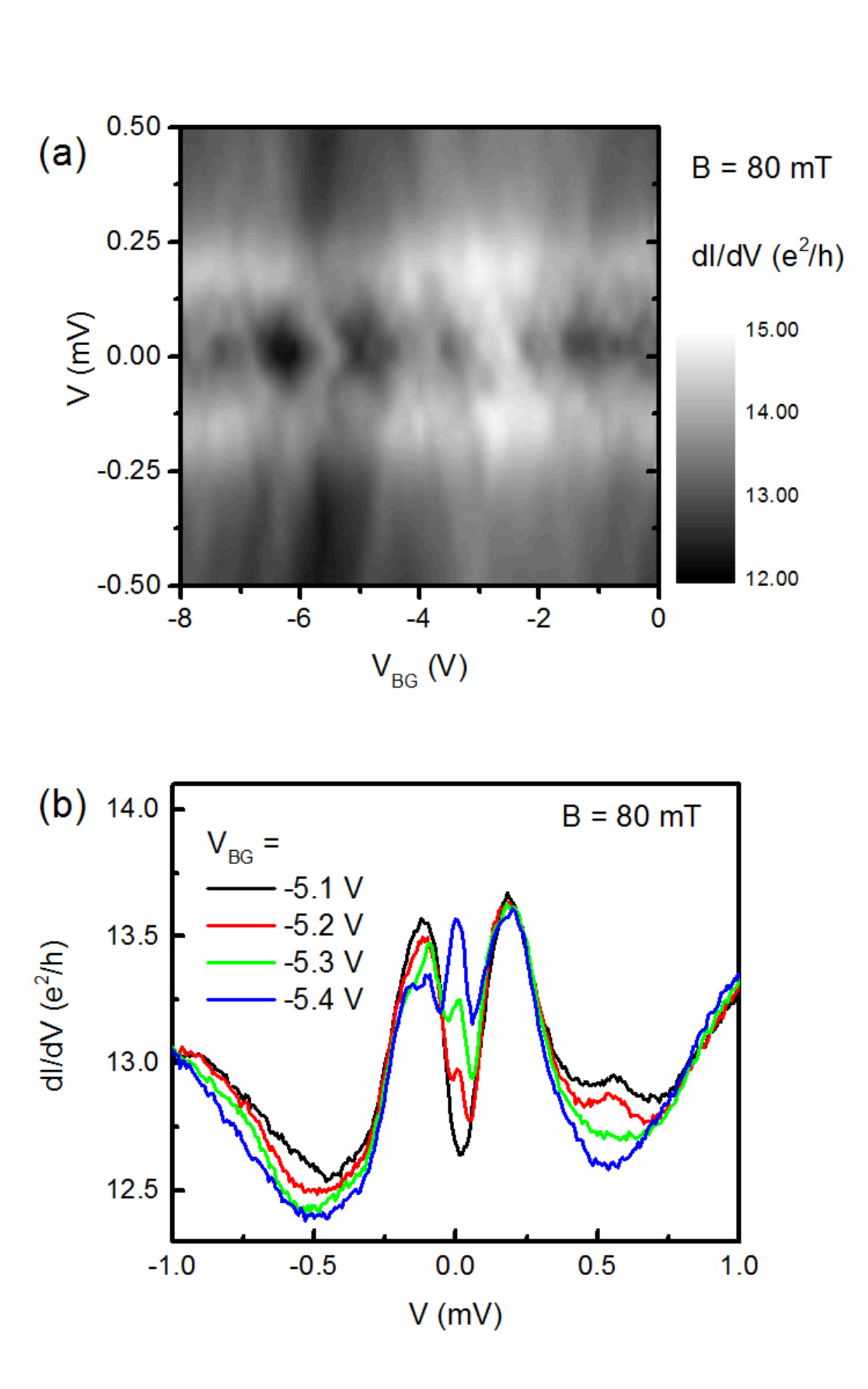}{(color online) (a) Color plot of $dI/dV$ vs source-drain bias $V$ and back gate bias$V_{BG}$ at B = 80 mT, with intricate low energy resonances.  (d) Example conductance traces at 80 mT, showing evolution of zero bias peak near $V_{BG} = -5.4$ V.}

\dograph{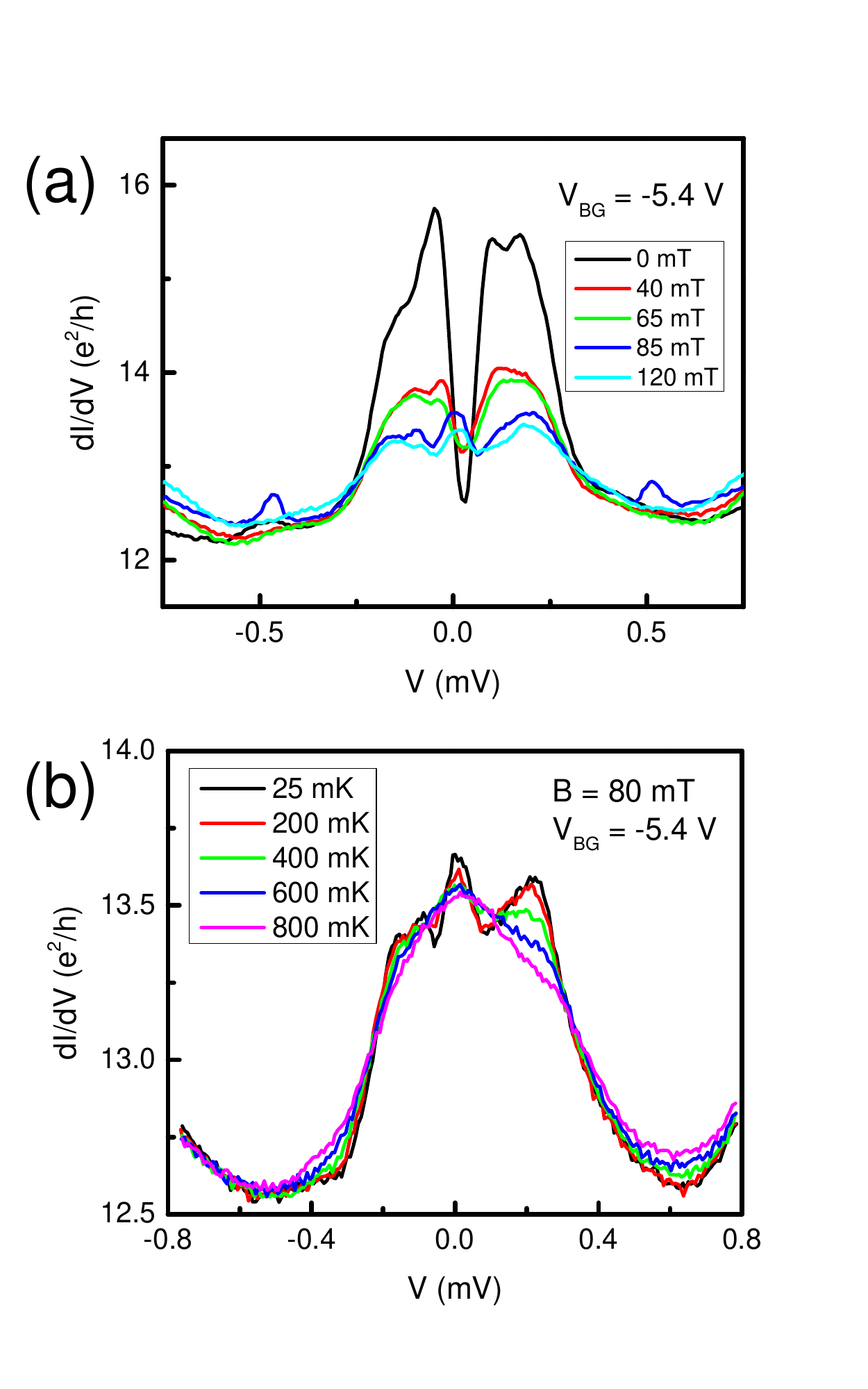}{(color online) (a) Magnetic field evolution of zero bias peak at $V_{BG} = -5.4$ V.  Peak appears beyond 65 mT and remains for up to $B= 120$ mT.  No such peak is present at zero field or when an in-plane field is applied perpendicular to current flow.  (b) Temperature dependence of zero bias peak at 80 mT.  Thermal fluctuations cause the peak to broaden and merge with background.}

These peaks are notable because no such features are seen in these two samples at zero magnetic field. They appear only when large fields are applied perpendicular to the TI surface and are accompanied by a collapse of the minigap governed by the Thouless energy (depicted in Fig.~\ref{fig:zba}a).  The low energy peaks do not appear when the TI flake is subjected to an in-plane magnetic field.  Thus, they share many of the properties of zero bias anomalies observed in spectroscopic studies of nanowires coupled to superconducting leads \cite{Mourik25052012, Deng2012, NPhysics2479, PhysRevLett.110.126406, PhysRevB.87.241401}.  While such studies were interpreted in terms of Majorana fermions, here we are more cautious.  For example, weak anti-localization can also generate low energy resonances due to coherence between electron-hole trajectories that repeatedly impinge upon the superconductor-semiconductor interface \cite{PhysRevLett.69.510}.  Although originally proposed to describe diffusive semiconductors, such trajectories can exist in ballistic cavities due to multiple reflections off the boundary opposite of the superconductor-semiconductor interface \cite{PhysRevB.64.224513}.  In this respect, this phenomenon is similar to the de Gennes-Saint-James resonances but is restricted to near zero energy.

It has been pointed out that weak anti-localization in superconductor-semiconductor devices could survive in the presence of broken time-reversal symmetry, leading to low energy anomalies that are nearly identical in appearance to those attributed to Majorana fermions \cite{NewJournalPhysics.14.125011}.  Such low energy peaks can emerge even in topologically trivial materials.  The persistence of de Gennes-Saint-James resonances and Fabry-Perot oscillations (Figs.~\ref{fig:fabryperotb}a and b) in our devices with magnetic field strongly suggests that the analogous trajectories for weak anti-localization can also remain phase-coherent.  This is in contrast to the previous assumption in the nanowire experiments that the large magnetic field would suppress weak anti-localization \cite{Mourik25052012, Deng2012, NPhysics2479, PhysRevLett.110.126406, PhysRevB.87.241401}.  We note that while Refs.~\cite{PhysRevLett.109.186802} and \cite{NatureNanotechnology.9.79} explored non-topological origins of zero bias anomalies in quantum dots coupled to superconducting leads, here we extend the experimental analysis to the limit of both ballistic transport and large coupling between the semiconductor segment and the metallic lead.

While the resonances in Fig.~\ref{fig:transport80mT}a might represent precursors to the low energy Andreev bound states that are predicted to form around the vortices of topological superconductors \cite{PhysRevLett.100.096407}, further work is required to confirm the existence of Majorana bound states.  Fortunately, our results demonstrate that it is feasible to construct Fabry-Perot interferometers on the surface of a topological insulator \cite{PhysRevLett.102.216403, PhysRevLett.102.216404, PhysRevLett.103.237001}.  By flowing current through chiral edge channels around the superconducting lead, one could conceivably probe the $Z_2$ interference of paths that enclose an even or odd number of vortices with Majorana fermions.  Such chiral edge channels will likely require inducing magnetization domains with ferromagnetic leads, which is beyond the subject of this present work.

\section{Conclusions}

We have presented transport measurements of the TI Bi$_2$Se$_3$ in contact with both a superconducting lead and a normal metal lead.  At low temperature and zero magnetic field, we find evidence of both Andreev reflection and the re-entrant resistance effect, confirming transparent contact between the TI and the superconductor.  By tuning the chemical potential with electrostatic gates, we observe clear Fabry-Perot oscillations with peak-to-peak amplitude $\approx 0.2$ $e^2/h$.  For source-drain biases below the energy gap of the superconducting lead, the Fabry-Perot oscillations double in frequency, demonstrating the emergence of de Gennes-Saint-James resonances.  In relatively large magnetic fields, a number of low-energy peaks are observed, which mimic the previously reported signatures of MBSs but strongly caution that they can also be explained in terms of weak anti-localization.  Our findings show an interplay between phase-coherent transport and Andreev reflection.  With a modified sample geometry, one can pursue interferometric searches of MBSs in TIs contacted with superconducting films.  The robustness of the Fabry-Perot oscillations despite multiple nanofabrication steps suggests that Bi$_2$Se$_3$ is a promising platform for more general studies of phase coherence in solid state systems.  Further study is needed to elucidate the role that topological surface states play in the Fabry-Perot oscillations.

\section{Acknowledgements}

We acknowledge helpful discussions with Rudro Biswas, Fiona Burnell, Liang Fu, Pouyan Ghaemi, Taylor Hughes, and Shu-Ping Lee.  C.K., A.D.K.F., and D.J.V.H.~acknowledge funding by Microsoft Project Q.  Y.S.H.~acknowledges support from National Science Foundation grant DMR-12-55607.  Device fabrication was carried out in the MRL Central Facilities (partially supported by the DOE under DE-FG02-07ER46453 and DE-FG02-07ER46471).

\bibliography{topological}

\begin{thebibliography}{58}%
\makeatletter
\providecommand \@ifxundefined [1]{%
 \@ifx{#1\undefined}
}%
\providecommand \@ifnum [1]{%
 \ifnum #1\expandafter \@firstoftwo
 \else \expandafter \@secondoftwo
 \fi
}%
\providecommand \@ifx [1]{%
 \ifx #1\expandafter \@firstoftwo
 \else \expandafter \@secondoftwo
 \fi
}%
\providecommand \natexlab [1]{#1}%
\providecommand \enquote  [1]{``#1''}%
\providecommand \bibnamefont  [1]{#1}%
\providecommand \bibfnamefont [1]{#1}%
\providecommand \citenamefont [1]{#1}%
\providecommand \href@noop [0]{\@secondoftwo}%
\providecommand \href [0]{\begingroup \@sanitize@url \@href}%
\providecommand \@href[1]{\@@startlink{#1}\@@href}%
\providecommand \@@href[1]{\endgroup#1\@@endlink}%
\providecommand \@sanitize@url [0]{\catcode `\\12\catcode `\$12\catcode
  `\&12\catcode `\#12\catcode `\^12\catcode `\_12\catcode `\%12\relax}%
\providecommand \@@startlink[1]{}%
\providecommand \@@endlink[0]{}%
\providecommand \url  [0]{\begingroup\@sanitize@url \@url }%
\providecommand \@url [1]{\endgroup\@href {#1}{\urlprefix }}%
\providecommand \urlprefix  [0]{URL }%
\providecommand \Eprint [0]{\href }%
\providecommand \doibase [0]{http://dx.doi.org/}%
\providecommand \selectlanguage [0]{\@gobble}%
\providecommand \bibinfo  [0]{\@secondoftwo}%
\providecommand \bibfield  [0]{\@secondoftwo}%
\providecommand \translation [1]{[#1]}%
\providecommand \BibitemOpen [0]{}%
\providecommand \bibitemStop [0]{}%
\providecommand \bibitemNoStop [0]{.\EOS\space}%
\providecommand \EOS [0]{\spacefactor3000\relax}%
\providecommand \BibitemShut  [1]{\csname bibitem#1\endcsname}%
\let\auto@bib@innerbib\@empty
\bibitem [{\citenamefont {Liang}\ \emph {et~al.}(2001)\citenamefont {Liang},
  \citenamefont {Bockrath}, \citenamefont {Bozovic}, \citenamefont {Hafner},
  \citenamefont {Tinkham},\ and\ \citenamefont {Park}}]{Nature.411.665}%
  \BibitemOpen
  \bibfield  {author} {\bibinfo {author} {\bibfnamefont {Wenjie}\ \bibnamefont
  {Liang}}, \bibinfo {author} {\bibfnamefont {Marc}\ \bibnamefont {Bockrath}},
  \bibinfo {author} {\bibfnamefont {Dolores}\ \bibnamefont {Bozovic}}, \bibinfo
  {author} {\bibfnamefont {Jason~H.}\ \bibnamefont {Hafner}}, \bibinfo {author}
  {\bibfnamefont {M.}~\bibnamefont {Tinkham}}, \ and\ \bibinfo {author}
  {\bibfnamefont {Hongkun}\ \bibnamefont {Park}},\ }\bibfield  {title}
  {\enquote {\bibinfo {title} {{Fabry - Perot interference in a nanotube
  electron waveguide}},}\ }\href {\doibase 10.1038/35079517;} {\bibfield
  {journal} {\bibinfo  {journal} {Nature}\ }\textbf {\bibinfo {volume} {411}},\
  \bibinfo {pages} {665} (\bibinfo {year} {2001})}\BibitemShut {NoStop}%
\bibitem [{\citenamefont {Kretinin}\ \emph {et~al.}(2010)\citenamefont
  {Kretinin}, \citenamefont {Popovitz-Biro}, \citenamefont {Mahalu},\ and\
  \citenamefont {Shtrikman}}]{NanoLett.10.3439}%
  \BibitemOpen
  \bibfield  {author} {\bibinfo {author} {\bibfnamefont {Andrey~V.}\
  \bibnamefont {Kretinin}}, \bibinfo {author} {\bibfnamefont {Ronit}\
  \bibnamefont {Popovitz-Biro}}, \bibinfo {author} {\bibfnamefont {Diana}\
  \bibnamefont {Mahalu}}, \ and\ \bibinfo {author} {\bibfnamefont {Hadas}\
  \bibnamefont {Shtrikman}},\ }\bibfield  {title} {\enquote {\bibinfo {title}
  {{Multimode Fabry-Pérot Conductance Oscillations in Suspended
  Stacking-Faults-Free InAs Nanowires}},}\ }\href {\doibase 10.1021/nl101522j}
  {\bibfield  {journal} {\bibinfo  {journal} {Nano Letters}\ }\textbf {\bibinfo
  {volume} {10}},\ \bibinfo {pages} {3439--3445} (\bibinfo {year}
  {2010})}\BibitemShut {NoStop}%
\bibitem [{\citenamefont {Miao}\ \emph {et~al.}(2007)\citenamefont {Miao},
  \citenamefont {Wijeratne}, \citenamefont {Zhang}, \citenamefont {Coskun},
  \citenamefont {Bao},\ and\ \citenamefont {Lau}}]{Miao14092007}%
  \BibitemOpen
  \bibfield  {author} {\bibinfo {author} {\bibfnamefont {F.}~\bibnamefont
  {Miao}}, \bibinfo {author} {\bibfnamefont {S.}~\bibnamefont {Wijeratne}},
  \bibinfo {author} {\bibfnamefont {Y.}~\bibnamefont {Zhang}}, \bibinfo
  {author} {\bibfnamefont {U.~C.}\ \bibnamefont {Coskun}}, \bibinfo {author}
  {\bibfnamefont {W.}~\bibnamefont {Bao}}, \ and\ \bibinfo {author}
  {\bibfnamefont {C.~N.}\ \bibnamefont {Lau}},\ }\bibfield  {title} {\enquote
  {\bibinfo {title} {{Phase-Coherent Transport in Graphene Quantum
  Billiards}},}\ }\href {\doibase 10.1126/science.1144359} {\bibfield
  {journal} {\bibinfo  {journal} {Science}\ }\textbf {\bibinfo {volume}
  {317}},\ \bibinfo {pages} {1530--1533} (\bibinfo {year} {2007})}\BibitemShut
  {NoStop}%
\bibitem [{\citenamefont {Hasan}\ and\ \citenamefont
  {Kane}(2010)}]{RevModPhys.82.3045}%
  \BibitemOpen
  \bibfield  {author} {\bibinfo {author} {\bibfnamefont {M.~Z.}\ \bibnamefont
  {Hasan}}\ and\ \bibinfo {author} {\bibfnamefont {C.~L.}\ \bibnamefont
  {Kane}},\ }\bibfield  {title} {\enquote {\bibinfo {title}
  {{\textit{Colloquium} : Topological insulators}},}\ }\href {\doibase
  10.1103/RevModPhys.82.3045} {\bibfield  {journal} {\bibinfo  {journal} {Rev.
  Mod. Phys.}\ }\textbf {\bibinfo {volume} {82}},\ \bibinfo {pages}
  {3045--3067} (\bibinfo {year} {2010})}\BibitemShut {NoStop}%
\bibitem [{\citenamefont {Nayak}\ \emph {et~al.}(2008)\citenamefont {Nayak},
  \citenamefont {Simon}, \citenamefont {Stern}, \citenamefont {Freedman},\ and\
  \citenamefont {Das~Sarma}}]{RevModPhys.80.1083}%
  \BibitemOpen
  \bibfield  {author} {\bibinfo {author} {\bibfnamefont {Chetan}\ \bibnamefont
  {Nayak}}, \bibinfo {author} {\bibfnamefont {Steven~H.}\ \bibnamefont
  {Simon}}, \bibinfo {author} {\bibfnamefont {Ady}\ \bibnamefont {Stern}},
  \bibinfo {author} {\bibfnamefont {Michael}\ \bibnamefont {Freedman}}, \ and\
  \bibinfo {author} {\bibfnamefont {Sankar}\ \bibnamefont {Das~Sarma}},\
  }\bibfield  {title} {\enquote {\bibinfo {title} {{Non-Abelian anyons and
  topological quantum computation}},}\ }\href {\doibase
  10.1103/RevModPhys.80.1083} {\bibfield  {journal} {\bibinfo  {journal} {Rev.
  Mod. Phys.}\ }\textbf {\bibinfo {volume} {80}},\ \bibinfo {pages}
  {1083--1159} (\bibinfo {year} {2008})}\BibitemShut {NoStop}%
\bibitem [{\citenamefont {Fu}\ and\ \citenamefont
  {Kane}(2008)}]{PhysRevLett.100.096407}%
  \BibitemOpen
  \bibfield  {author} {\bibinfo {author} {\bibfnamefont {Liang}\ \bibnamefont
  {Fu}}\ and\ \bibinfo {author} {\bibfnamefont {C.~L.}\ \bibnamefont {Kane}},\
  }\bibfield  {title} {\enquote {\bibinfo {title} {{Superconducting Proximity
  Effect and Majorana Fermions at the Surface of a Topological Insulator}},}\
  }\href {\doibase 10.1103/PhysRevLett.100.096407} {\bibfield  {journal}
  {\bibinfo  {journal} {Phys. Rev. Lett.}\ }\textbf {\bibinfo {volume} {100}},\
  \bibinfo {pages} {096407} (\bibinfo {year} {2008})}\BibitemShut {NoStop}%
\bibitem [{\citenamefont {Alicea}(2012)}]{Alicea2012}%
  \BibitemOpen
  \bibfield  {author} {\bibinfo {author} {\bibfnamefont {J.}~\bibnamefont
  {Alicea}},\ }\bibfield  {title} {\enquote {\bibinfo {title} {{New directions
  in the pursuit of Majorana fermions in solid state systems}},}\ }\href@noop
  {} {\bibfield  {journal} {\bibinfo  {journal} {Rep. Prog. Phys.}\ }\textbf
  {\bibinfo {volume} {75}},\ \bibinfo {pages} {076501} (\bibinfo {year}
  {2012})}\BibitemShut {NoStop}%
\bibitem [{\citenamefont {Beenakker}(2013)}]{Beenakker2013}%
  \BibitemOpen
  \bibfield  {author} {\bibinfo {author} {\bibfnamefont {C.W.J.}\ \bibnamefont
  {Beenakker}},\ }\bibfield  {title} {\enquote {\bibinfo {title} {{Search for
  Majorana fermions in superconductors}},}\ }\href {\doibase
  10.1146/annurev-conmatphys-030212-184337} {\bibfield  {journal} {\bibinfo
  {journal} {Annu. Rev. Con. Mat. Phys.}\ }\textbf {\bibinfo {volume} {4}},\
  \bibinfo {pages} {113} (\bibinfo {year} {2013})}\BibitemShut {NoStop}%
\bibitem [{\citenamefont {Fu}\ and\ \citenamefont
  {Kane}(2009)}]{PhysRevLett.102.216403}%
  \BibitemOpen
  \bibfield  {author} {\bibinfo {author} {\bibfnamefont {Liang}\ \bibnamefont
  {Fu}}\ and\ \bibinfo {author} {\bibfnamefont {C.~L.}\ \bibnamefont {Kane}},\
  }\bibfield  {title} {\enquote {\bibinfo {title} {{Probing Neutral Majorana
  Fermion Edge Modes with Charge Transport}},}\ }\href {\doibase
  10.1103/PhysRevLett.102.216403} {\bibfield  {journal} {\bibinfo  {journal}
  {Phys. Rev. Lett.}\ }\textbf {\bibinfo {volume} {102}},\ \bibinfo {pages}
  {216403} (\bibinfo {year} {2009})}\BibitemShut {NoStop}%
\bibitem [{\citenamefont {Akhmerov}\ \emph {et~al.}(2009)\citenamefont
  {Akhmerov}, \citenamefont {Nilsson},\ and\ \citenamefont
  {Beenakker}}]{PhysRevLett.102.216404}%
  \BibitemOpen
  \bibfield  {author} {\bibinfo {author} {\bibfnamefont {A.~R.}\ \bibnamefont
  {Akhmerov}}, \bibinfo {author} {\bibfnamefont {Johan}\ \bibnamefont
  {Nilsson}}, \ and\ \bibinfo {author} {\bibfnamefont {C.~W.~J.}\ \bibnamefont
  {Beenakker}},\ }\bibfield  {title} {\enquote {\bibinfo {title} {{Electrically
  Detected Interferometry of Majorana Fermions in a Topological Insulator}},}\
  }\href {\doibase 10.1103/PhysRevLett.102.216404} {\bibfield  {journal}
  {\bibinfo  {journal} {Phys. Rev. Lett.}\ }\textbf {\bibinfo {volume} {102}},\
  \bibinfo {pages} {216404} (\bibinfo {year} {2009})}\BibitemShut {NoStop}%
\bibitem [{\citenamefont {Law}\ \emph {et~al.}(2009)\citenamefont {Law},
  \citenamefont {Lee},\ and\ \citenamefont {Ng}}]{PhysRevLett.103.237001}%
  \BibitemOpen
  \bibfield  {author} {\bibinfo {author} {\bibfnamefont {K.~T.}\ \bibnamefont
  {Law}}, \bibinfo {author} {\bibfnamefont {Patrick~A.}\ \bibnamefont {Lee}}, \
  and\ \bibinfo {author} {\bibfnamefont {T.~K.}\ \bibnamefont {Ng}},\
  }\bibfield  {title} {\enquote {\bibinfo {title} {{Majorana Fermion Induced
  Resonant Andreev Reflection}},}\ }\href {\doibase
  10.1103/PhysRevLett.103.237001} {\bibfield  {journal} {\bibinfo  {journal}
  {Phys. Rev. Lett.}\ }\textbf {\bibinfo {volume} {103}},\ \bibinfo {pages}
  {237001} (\bibinfo {year} {2009})}\BibitemShut {NoStop}%
\bibitem [{\citenamefont {Zhang}\ \emph {et~al.}(2011)\citenamefont {Zhang},
  \citenamefont {Wang}, \citenamefont {DaSilva}, \citenamefont {Lee},
  \citenamefont {Gutierrez}, \citenamefont {Chan}, \citenamefont {Jain},\ and\
  \citenamefont {Samarth}}]{PhysRevB.84.165120}%
  \BibitemOpen
  \bibfield  {author} {\bibinfo {author} {\bibfnamefont {Duming}\ \bibnamefont
  {Zhang}}, \bibinfo {author} {\bibfnamefont {Jian}\ \bibnamefont {Wang}},
  \bibinfo {author} {\bibfnamefont {Ashley~M.}\ \bibnamefont {DaSilva}},
  \bibinfo {author} {\bibfnamefont {Joon~Sue}\ \bibnamefont {Lee}}, \bibinfo
  {author} {\bibfnamefont {Humberto~R.}\ \bibnamefont {Gutierrez}}, \bibinfo
  {author} {\bibfnamefont {Moses H.~W.}\ \bibnamefont {Chan}}, \bibinfo
  {author} {\bibfnamefont {Jainendra}\ \bibnamefont {Jain}}, \ and\ \bibinfo
  {author} {\bibfnamefont {Nitin}\ \bibnamefont {Samarth}},\ }\bibfield
  {title} {\enquote {\bibinfo {title} {{Superconducting proximity effect and
  possible evidence for Pearl vortices in a candidate topological
  insulator}},}\ }\href {\doibase 10.1103/PhysRevB.84.165120} {\bibfield
  {journal} {\bibinfo  {journal} {Phys. Rev. B}\ }\textbf {\bibinfo {volume}
  {84}},\ \bibinfo {pages} {165120} (\bibinfo {year} {2011})}\BibitemShut
  {NoStop}%
\bibitem [{\citenamefont {Sac\'{e}p\'{e}}\ \emph {et~al.}(2011)\citenamefont
  {Sac\'{e}p\'{e}}, \citenamefont {Oostinga}, \citenamefont {Li}, \citenamefont
  {Ubaldini}, \citenamefont {Couto}, \citenamefont {Giannini},\ and\
  \citenamefont {Morpurgo}}]{NatCommun.2.575}%
  \BibitemOpen
  \bibfield  {author} {\bibinfo {author} {\bibfnamefont {Benjamin}\
  \bibnamefont {Sac\'{e}p\'{e}}}, \bibinfo {author} {\bibfnamefont {Jeroen~B.}\
  \bibnamefont {Oostinga}}, \bibinfo {author} {\bibfnamefont {Jian}\
  \bibnamefont {Li}}, \bibinfo {author} {\bibfnamefont {Alberto}\ \bibnamefont
  {Ubaldini}}, \bibinfo {author} {\bibfnamefont {Nuno~J.G.}\ \bibnamefont
  {Couto}}, \bibinfo {author} {\bibfnamefont {Enrico}\ \bibnamefont
  {Giannini}}, \ and\ \bibinfo {author} {\bibfnamefont {Alberto~F.}\
  \bibnamefont {Morpurgo}},\ }\bibfield  {title} {\enquote {\bibinfo {title}
  {{Gate-tuned normal and superconducting transport at the surface of a
  topological insulator}},}\ }\href {http://dx.doi.org/10.1038/ncomms1586}
  {\bibfield  {journal} {\bibinfo  {journal} {Nat. Commun.}\ }\textbf {\bibinfo
  {volume} {2}},\ \bibinfo {pages} {575} (\bibinfo {year} {2011})}\BibitemShut
  {NoStop}%
\bibitem [{\citenamefont {Qu}\ \emph {et~al.}(2012)\citenamefont {Qu},
  \citenamefont {Yang}, \citenamefont {Shen}, \citenamefont {Ding},
  \citenamefont {Chen}, \citenamefont {Ji}, \citenamefont {Liu}, \citenamefont
  {Fan}, \citenamefont {Jing}, \citenamefont {Yang},\ and\ \citenamefont
  {Lu}}]{SciRep.2.339}%
  \BibitemOpen
  \bibfield  {author} {\bibinfo {author} {\bibfnamefont {Fanming}\ \bibnamefont
  {Qu}}, \bibinfo {author} {\bibfnamefont {Fan}\ \bibnamefont {Yang}}, \bibinfo
  {author} {\bibfnamefont {Jie}\ \bibnamefont {Shen}}, \bibinfo {author}
  {\bibfnamefont {Yue}\ \bibnamefont {Ding}}, \bibinfo {author} {\bibfnamefont
  {Jun}\ \bibnamefont {Chen}}, \bibinfo {author} {\bibfnamefont {Zhongqing}\
  \bibnamefont {Ji}}, \bibinfo {author} {\bibfnamefont {Guangtong}\
  \bibnamefont {Liu}}, \bibinfo {author} {\bibfnamefont {Jie}\ \bibnamefont
  {Fan}}, \bibinfo {author} {\bibfnamefont {Xiunian}\ \bibnamefont {Jing}},
  \bibinfo {author} {\bibfnamefont {Changli}\ \bibnamefont {Yang}}, \ and\
  \bibinfo {author} {\bibfnamefont {Li}~\bibnamefont {Lu}},\ }\bibfield
  {title} {\enquote {\bibinfo {title} {{Strong Superconducting Proximity Effect
  in Pb-Bi2Te3 Hybrid Structures}},}\ }\href
  {http://dx.doi.org/10.1038/srep00339} {\bibfield  {journal} {\bibinfo
  {journal} {Sci. Rep.}\ }\textbf {\bibinfo {volume} {2}},\ \bibinfo {pages}
  {339} (\bibinfo {year} {2012})}\BibitemShut {NoStop}%
\bibitem [{\citenamefont {Veldhorst}\ \emph {et~al.}(2012)\citenamefont
  {Veldhorst}, \citenamefont {Snelder}, \citenamefont {Hoek}, \citenamefont
  {Gang}, \citenamefont {Guduru}, \citenamefont {Wang}, \citenamefont
  {Zeitler}, \citenamefont {van~der Wiel}, \citenamefont {Golubov},
  \citenamefont {Hilgenkamp},\ and\ \citenamefont {Brinkman}}]{NatMat.11.417}%
  \BibitemOpen
  \bibfield  {author} {\bibinfo {author} {\bibfnamefont {M.}~\bibnamefont
  {Veldhorst}}, \bibinfo {author} {\bibfnamefont {M.}~\bibnamefont {Snelder}},
  \bibinfo {author} {\bibfnamefont {M.}~\bibnamefont {Hoek}}, \bibinfo {author}
  {\bibfnamefont {T.}~\bibnamefont {Gang}}, \bibinfo {author} {\bibfnamefont
  {V.~K.}\ \bibnamefont {Guduru}}, \bibinfo {author} {\bibfnamefont {X.~L.}\
  \bibnamefont {Wang}}, \bibinfo {author} {\bibfnamefont {U.}~\bibnamefont
  {Zeitler}}, \bibinfo {author} {\bibfnamefont {W.~G.}\ \bibnamefont {van~der
  Wiel}}, \bibinfo {author} {\bibfnamefont {A.~A.}\ \bibnamefont {Golubov}},
  \bibinfo {author} {\bibfnamefont {H.}~\bibnamefont {Hilgenkamp}}, \ and\
  \bibinfo {author} {\bibfnamefont {A.}~\bibnamefont {Brinkman}},\ }\bibfield
  {title} {\enquote {\bibinfo {title} {{Josephson supercurrent through a
  topological insulator surface state}},}\ }\href {dx.doi.org/10.1038/nmat3255}
  {\bibfield  {journal} {\bibinfo  {journal} {Nat. Mater.}\ }\textbf {\bibinfo
  {volume} {11}},\ \bibinfo {pages} {417} (\bibinfo {year} {2012})}\BibitemShut
  {NoStop}%
\bibitem [{\citenamefont {Cho}\ \emph {et~al.}(2013)\citenamefont {Cho},
  \citenamefont {Dellabetta}, \citenamefont {Yang}, \citenamefont {Schneeloch},
  \citenamefont {Xu}, \citenamefont {Valla}, \citenamefont {Gu}, \citenamefont
  {Gilbert},\ and\ \citenamefont {Mason}}]{NatCommun.4.1689}%
  \BibitemOpen
  \bibfield  {author} {\bibinfo {author} {\bibfnamefont {Sungjae}\ \bibnamefont
  {Cho}}, \bibinfo {author} {\bibfnamefont {Brian}\ \bibnamefont {Dellabetta}},
  \bibinfo {author} {\bibfnamefont {Alina}\ \bibnamefont {Yang}}, \bibinfo
  {author} {\bibfnamefont {John}\ \bibnamefont {Schneeloch}}, \bibinfo {author}
  {\bibfnamefont {Zhijun}\ \bibnamefont {Xu}}, \bibinfo {author} {\bibfnamefont
  {Tonica}\ \bibnamefont {Valla}}, \bibinfo {author} {\bibfnamefont {Genda}\
  \bibnamefont {Gu}}, \bibinfo {author} {\bibfnamefont {Matthew~J.}\
  \bibnamefont {Gilbert}}, \ and\ \bibinfo {author} {\bibfnamefont {Nadya}\
  \bibnamefont {Mason}},\ }\bibfield  {title} {\enquote {\bibinfo {title}
  {{Symmetry protected Josephson supercurrents in three-dimensional topological
  insulators}},}\ }\href {http://dx.doi.org/10.1038/ncomms2701} {\bibfield
  {journal} {\bibinfo  {journal} {Nat. Commun.}\ }\textbf {\bibinfo {volume}
  {4}},\ \bibinfo {pages} {1689} (\bibinfo {year} {2013})}\BibitemShut
  {NoStop}%
\bibitem [{\citenamefont {Oostinga}\ \emph {et~al.}(2013)\citenamefont
  {Oostinga}, \citenamefont {Maier}, \citenamefont {Sch\"uffelgen},
  \citenamefont {Knott}, \citenamefont {Ames}, \citenamefont {Br\"une},
  \citenamefont {Tkachov}, \citenamefont {Buhmann},\ and\ \citenamefont
  {Molenkamp}}]{PhysRevX.3.021007}%
  \BibitemOpen
  \bibfield  {author} {\bibinfo {author} {\bibfnamefont {Jeroen~B.}\
  \bibnamefont {Oostinga}}, \bibinfo {author} {\bibfnamefont {Luis}\
  \bibnamefont {Maier}}, \bibinfo {author} {\bibfnamefont {Peter}\ \bibnamefont
  {Sch\"uffelgen}}, \bibinfo {author} {\bibfnamefont {Daniel}\ \bibnamefont
  {Knott}}, \bibinfo {author} {\bibfnamefont {Christopher}\ \bibnamefont
  {Ames}}, \bibinfo {author} {\bibfnamefont {Christoph}\ \bibnamefont
  {Br\"une}}, \bibinfo {author} {\bibfnamefont {Grigory}\ \bibnamefont
  {Tkachov}}, \bibinfo {author} {\bibfnamefont {Hartmut}\ \bibnamefont
  {Buhmann}}, \ and\ \bibinfo {author} {\bibfnamefont {Laurens~W.}\
  \bibnamefont {Molenkamp}},\ }\bibfield  {title} {\enquote {\bibinfo {title}
  {{Josephson Supercurrent through the Topological Surface States of Strained
  Bulk HgTe}},}\ }\href {\doibase 10.1103/PhysRevX.3.021007} {\bibfield
  {journal} {\bibinfo  {journal} {Phys. Rev. X}\ }\textbf {\bibinfo {volume}
  {3}},\ \bibinfo {pages} {021007} (\bibinfo {year} {2013})}\BibitemShut
  {NoStop}%
\bibitem [{\citenamefont {Orlyanchik}\ \emph {et~al.}(2013)\citenamefont
  {Orlyanchik}, \citenamefont {Stehno}, \citenamefont {Nugroho}, \citenamefont
  {Ghaemi}, \citenamefont {Brahlek}, \citenamefont {Koirala}, \citenamefont
  {Oh},\ and\ \citenamefont {Van~Harlingen}}]{Orlyanchik2013}%
  \BibitemOpen
  \bibfield  {author} {\bibinfo {author} {\bibfnamefont {Vladimir}\
  \bibnamefont {Orlyanchik}}, \bibinfo {author} {\bibfnamefont {Martin~P.}\
  \bibnamefont {Stehno}}, \bibinfo {author} {\bibfnamefont {Christopher~D.}\
  \bibnamefont {Nugroho}}, \bibinfo {author} {\bibfnamefont {Pouyan}\
  \bibnamefont {Ghaemi}}, \bibinfo {author} {\bibfnamefont {Matthew}\
  \bibnamefont {Brahlek}}, \bibinfo {author} {\bibfnamefont {Nikesh}\
  \bibnamefont {Koirala}}, \bibinfo {author} {\bibfnamefont {Seongshik}\
  \bibnamefont {Oh}}, \ and\ \bibinfo {author} {\bibfnamefont {Dale~J.}\
  \bibnamefont {Van~Harlingen}},\ }\bibfield  {title} {\enquote {\bibinfo
  {title} {{Signature of a topological phase transition in the Josephson
  supercurrent through a topological insulator}},}\ }\href
  {http://arxiv.org/abs/1309.0163} {\bibfield  {journal} {\bibinfo  {journal}
  {arXiv:1309.0163}\ } (\bibinfo {year} {2013})}\BibitemShut {NoStop}%
\bibitem [{\citenamefont {Kurter}\ \emph {et~al.}(2013)\citenamefont {Kurter},
  \citenamefont {Finck}, \citenamefont {Hor},\ and\ \citenamefont
  {Van~Harlingen}}]{Kurter2013}%
  \BibitemOpen
  \bibfield  {author} {\bibinfo {author} {\bibfnamefont {Cihan}\ \bibnamefont
  {Kurter}}, \bibinfo {author} {\bibfnamefont {Aaron D.~K.}\ \bibnamefont
  {Finck}}, \bibinfo {author} {\bibfnamefont {Yew~San}\ \bibnamefont {Hor}}, \
  and\ \bibinfo {author} {\bibfnamefont {Dale~J.}\ \bibnamefont
  {Van~Harlingen}},\ }\bibfield  {title} {\enquote {\bibinfo {title} {{Evidence
  for an anomalous current-phase relation of a dc SQUID with tunable
  topological junctions}},}\ }\href {http://arxiv.org/abs/1307.7764} {\bibfield
   {journal} {\bibinfo  {journal} {arXiv:1307.7764}\ } (\bibinfo {year}
  {2013})}\BibitemShut {NoStop}%
\bibitem [{\citenamefont {Galletti}\ \emph {et~al.}(2014)\citenamefont
  {Galletti}, \citenamefont {Charpentier}, \citenamefont {Iavarone},
  \citenamefont {Lucignano}, \citenamefont {Massarotti}, \citenamefont
  {Arpaia}, \citenamefont {Suzuki}, \citenamefont {Kadowaki}, \citenamefont
  {Bauch}, \citenamefont {Tagliacozzo}, \citenamefont {Tafuri},\ and\
  \citenamefont {Lombardi}}]{PhysRevB.89.134512}%
  \BibitemOpen
  \bibfield  {author} {\bibinfo {author} {\bibfnamefont {L.}~\bibnamefont
  {Galletti}}, \bibinfo {author} {\bibfnamefont {S.}~\bibnamefont
  {Charpentier}}, \bibinfo {author} {\bibfnamefont {M.}~\bibnamefont
  {Iavarone}}, \bibinfo {author} {\bibfnamefont {P.}~\bibnamefont {Lucignano}},
  \bibinfo {author} {\bibfnamefont {D.}~\bibnamefont {Massarotti}}, \bibinfo
  {author} {\bibfnamefont {R.}~\bibnamefont {Arpaia}}, \bibinfo {author}
  {\bibfnamefont {Y.}~\bibnamefont {Suzuki}}, \bibinfo {author} {\bibfnamefont
  {K.}~\bibnamefont {Kadowaki}}, \bibinfo {author} {\bibfnamefont
  {T.}~\bibnamefont {Bauch}}, \bibinfo {author} {\bibfnamefont
  {A.}~\bibnamefont {Tagliacozzo}}, \bibinfo {author} {\bibfnamefont
  {F.}~\bibnamefont {Tafuri}}, \ and\ \bibinfo {author} {\bibfnamefont
  {F.}~\bibnamefont {Lombardi}},\ }\bibfield  {title} {\enquote {\bibinfo
  {title} {{Influence of topological edge states on the properties of
  Al/Bi$_2$Se$_3$/Al hybrid Josephson devices}},}\ }\href {\doibase
  10.1103/PhysRevB.89.134512} {\bibfield  {journal} {\bibinfo  {journal} {Phys.
  Rev. B}\ }\textbf {\bibinfo {volume} {89}},\ \bibinfo {pages} {134512}
  (\bibinfo {year} {2014})}\BibitemShut {NoStop}%
\bibitem [{\citenamefont {Kurter}\ \emph {et~al.}(2014)\citenamefont {Kurter},
  \citenamefont {Finck}, \citenamefont {Ghaemi}, \citenamefont {Hor},\ and\
  \citenamefont {Van~Harlingen}}]{PhysRevB.90.014501}%
  \BibitemOpen
  \bibfield  {author} {\bibinfo {author} {\bibfnamefont {C.}~\bibnamefont
  {Kurter}}, \bibinfo {author} {\bibfnamefont {A.~D.~K.}\ \bibnamefont
  {Finck}}, \bibinfo {author} {\bibfnamefont {P.}~\bibnamefont {Ghaemi}},
  \bibinfo {author} {\bibfnamefont {Y.~S.}\ \bibnamefont {Hor}}, \ and\
  \bibinfo {author} {\bibfnamefont {D.~J.}\ \bibnamefont {Van~Harlingen}},\
  }\bibfield  {title} {\enquote {\bibinfo {title} {Dynamical gate-tunable
  supercurrents in topological josephson junctions},}\ }\href {\doibase
  10.1103/PhysRevB.90.014501} {\bibfield  {journal} {\bibinfo  {journal} {Phys.
  Rev. B}\ }\textbf {\bibinfo {volume} {90}},\ \bibinfo {pages} {014501}
  (\bibinfo {year} {2014})}\BibitemShut {NoStop}%
\bibitem [{\citenamefont {de~Gennes}\ and\ \citenamefont
  {Saint-James}(1963)}]{PhysLett.4.151}%
  \BibitemOpen
  \bibfield  {author} {\bibinfo {author} {\bibfnamefont {P.G.}\ \bibnamefont
  {de~Gennes}}\ and\ \bibinfo {author} {\bibfnamefont {D.}~\bibnamefont
  {Saint-James}},\ }\bibfield  {title} {\enquote {\bibinfo {title} {{Elementary
  excitations in the vicinity of a normal metal-superconducting metal
  contact}},}\ }\href {http://dx.doi.org/10.1016/0031-9163(63)90148-3}
  {\bibfield  {journal} {\bibinfo  {journal} {Phys. Lett.}\ }\textbf {\bibinfo
  {volume} {4}},\ \bibinfo {pages} {151} (\bibinfo {year} {1963})}\BibitemShut
  {NoStop}%
\bibitem [{\citenamefont {Rowell}\ and\ \citenamefont
  {McMillan}(1966)}]{PhysRevLett.16.453}%
  \BibitemOpen
  \bibfield  {author} {\bibinfo {author} {\bibfnamefont {J.~M.}\ \bibnamefont
  {Rowell}}\ and\ \bibinfo {author} {\bibfnamefont {W.~L.}\ \bibnamefont
  {McMillan}},\ }\bibfield  {title} {\enquote {\bibinfo {title} {{Electron
  Interference in a Normal Metal Induced by Superconducting Contracts}},}\
  }\href {\doibase 10.1103/PhysRevLett.16.453} {\bibfield  {journal} {\bibinfo
  {journal} {Phys. Rev. Lett.}\ }\textbf {\bibinfo {volume} {16}},\ \bibinfo
  {pages} {453--456} (\bibinfo {year} {1966})}\BibitemShut {NoStop}%
\bibitem [{\citenamefont {Mourik}\ \emph {et~al.}(2012)\citenamefont {Mourik},
  \citenamefont {Zuo}, \citenamefont {Frolov}, \citenamefont {Plissard},
  \citenamefont {Bakkers},\ and\ \citenamefont {Kouwenhoven}}]{Mourik25052012}%
  \BibitemOpen
  \bibfield  {author} {\bibinfo {author} {\bibfnamefont {V.}~\bibnamefont
  {Mourik}}, \bibinfo {author} {\bibfnamefont {K.}~\bibnamefont {Zuo}},
  \bibinfo {author} {\bibfnamefont {S.M.}\ \bibnamefont {Frolov}}, \bibinfo
  {author} {\bibfnamefont {S.R.}\ \bibnamefont {Plissard}}, \bibinfo {author}
  {\bibfnamefont {E.P.A.M.}\ \bibnamefont {Bakkers}}, \ and\ \bibinfo {author}
  {\bibfnamefont {L.P.}\ \bibnamefont {Kouwenhoven}},\ }\bibfield  {title}
  {\enquote {\bibinfo {title} {{Signatures of Majorana Fermions in Hybrid
  Superconductor-Semiconductor Nanowire Devices}},}\ }\href {\doibase
  10.1126/science.1222360} {\bibfield  {journal} {\bibinfo  {journal}
  {Science}\ }\textbf {\bibinfo {volume} {336}},\ \bibinfo {pages} {1003--1007}
  (\bibinfo {year} {2012})}\BibitemShut {NoStop}%
\bibitem [{\citenamefont {Deng}\ \emph {et~al.}(2012)\citenamefont {Deng},
  \citenamefont {Yu}, \citenamefont {Huang}, \citenamefont {Larsson},
  \citenamefont {Caroff},\ and\ \citenamefont {Xu}}]{Deng2012}%
  \BibitemOpen
  \bibfield  {author} {\bibinfo {author} {\bibfnamefont {M.~T.}\ \bibnamefont
  {Deng}}, \bibinfo {author} {\bibfnamefont {C.~L.}\ \bibnamefont {Yu}},
  \bibinfo {author} {\bibfnamefont {G.~Y.}\ \bibnamefont {Huang}}, \bibinfo
  {author} {\bibfnamefont {M.}~\bibnamefont {Larsson}}, \bibinfo {author}
  {\bibfnamefont {P.}~\bibnamefont {Caroff}}, \ and\ \bibinfo {author}
  {\bibfnamefont {H.~Q.}\ \bibnamefont {Xu}},\ }\bibfield  {title} {\enquote
  {\bibinfo {title} {{Anomalous Zero-Bias Conductance Peak in a Nb-InSb
  Nanowire-Nb Hybrid Device}},}\ }\href {\doibase 10.1021/nl303758w} {\bibfield
   {journal} {\bibinfo  {journal} {Nano Letters}\ }\textbf {\bibinfo {volume}
  {12}},\ \bibinfo {pages} {6414} (\bibinfo {year} {2012})}\BibitemShut
  {NoStop}%
\bibitem [{\citenamefont {Das}\ \emph {et~al.}(2012)\citenamefont {Das},
  \citenamefont {Ronen}, \citenamefont {Most}, \citenamefont {Oreg},
  \citenamefont {Heiblum},\ and\ \citenamefont {Shtrikman}}]{NPhysics2479}%
  \BibitemOpen
  \bibfield  {author} {\bibinfo {author} {\bibfnamefont {A.}~\bibnamefont
  {Das}}, \bibinfo {author} {\bibfnamefont {Y.}~\bibnamefont {Ronen}}, \bibinfo
  {author} {\bibfnamefont {Y.}~\bibnamefont {Most}}, \bibinfo {author}
  {\bibfnamefont {Y.}~\bibnamefont {Oreg}}, \bibinfo {author} {\bibfnamefont
  {M.}~\bibnamefont {Heiblum}}, \ and\ \bibinfo {author} {\bibfnamefont
  {H.}~\bibnamefont {Shtrikman}},\ }\bibfield  {title} {\enquote {\bibinfo
  {title} {{Zero-bias peaks and splitting in an Al-InAs nanowire topological
  superconductor as a signature of Majorana fermions}},}\ }\href {\doibase
  10.1038/nphys2479} {\bibfield  {journal} {\bibinfo  {journal} {Nature Phys.}\
  }\textbf {\bibinfo {volume} {8}},\ \bibinfo {pages} {887} (\bibinfo {year}
  {2012})}\BibitemShut {NoStop}%
\bibitem [{\citenamefont {Finck}\ \emph {et~al.}(2013)\citenamefont {Finck},
  \citenamefont {Van~Harlingen}, \citenamefont {Mohseni}, \citenamefont
  {Jung},\ and\ \citenamefont {Li}}]{PhysRevLett.110.126406}%
  \BibitemOpen
  \bibfield  {author} {\bibinfo {author} {\bibfnamefont {A.~D.~K.}\
  \bibnamefont {Finck}}, \bibinfo {author} {\bibfnamefont {D.~J.}\ \bibnamefont
  {Van~Harlingen}}, \bibinfo {author} {\bibfnamefont {P.~K.}\ \bibnamefont
  {Mohseni}}, \bibinfo {author} {\bibfnamefont {K.}~\bibnamefont {Jung}}, \
  and\ \bibinfo {author} {\bibfnamefont {X.}~\bibnamefont {Li}},\ }\bibfield
  {title} {\enquote {\bibinfo {title} {{Anomalous Modulation of a Zero-Bias
  Peak in a Hybrid Nanowire-Superconductor Device}},}\ }\href {\doibase
  10.1103/PhysRevLett.110.126406} {\bibfield  {journal} {\bibinfo  {journal}
  {Phys. Rev. Lett.}\ }\textbf {\bibinfo {volume} {110}},\ \bibinfo {pages}
  {126406} (\bibinfo {year} {2013})}\BibitemShut {NoStop}%
\bibitem [{\citenamefont {Churchill}\ \emph {et~al.}(2013)\citenamefont
  {Churchill}, \citenamefont {Fatemi}, \citenamefont {Grove-Rasmussen},
  \citenamefont {Deng}, \citenamefont {Caroff}, \citenamefont {Xu},\ and\
  \citenamefont {Marcus}}]{PhysRevB.87.241401}%
  \BibitemOpen
  \bibfield  {author} {\bibinfo {author} {\bibfnamefont {H.~O.~H.}\
  \bibnamefont {Churchill}}, \bibinfo {author} {\bibfnamefont {V.}~\bibnamefont
  {Fatemi}}, \bibinfo {author} {\bibfnamefont {K.}~\bibnamefont
  {Grove-Rasmussen}}, \bibinfo {author} {\bibfnamefont {M.~T.}\ \bibnamefont
  {Deng}}, \bibinfo {author} {\bibfnamefont {P.}~\bibnamefont {Caroff}},
  \bibinfo {author} {\bibfnamefont {H.~Q.}\ \bibnamefont {Xu}}, \ and\ \bibinfo
  {author} {\bibfnamefont {C.~M.}\ \bibnamefont {Marcus}},\ }\bibfield  {title}
  {\enquote {\bibinfo {title} {{Superconductor-nanowire devices from tunneling
  to the multichannel regime: Zero-bias oscillations and magnetoconductance
  crossover}},}\ }\href {\doibase 10.1103/PhysRevB.87.241401} {\bibfield
  {journal} {\bibinfo  {journal} {Phys. Rev. B}\ }\textbf {\bibinfo {volume}
  {87}},\ \bibinfo {pages} {241401} (\bibinfo {year} {2013})}\BibitemShut
  {NoStop}%
\bibitem [{\citenamefont {Pikulin}\ \emph {et~al.}(2012)\citenamefont
  {Pikulin}, \citenamefont {Dahlhaus}, \citenamefont {Wimmer}, \citenamefont
  {Schomerus},\ and\ \citenamefont {Beenakker}}]{NewJournalPhysics.14.125011}%
  \BibitemOpen
  \bibfield  {author} {\bibinfo {author} {\bibfnamefont {D.I.}\ \bibnamefont
  {Pikulin}}, \bibinfo {author} {\bibfnamefont {J.P.}\ \bibnamefont
  {Dahlhaus}}, \bibinfo {author} {\bibfnamefont {M.}~\bibnamefont {Wimmer}},
  \bibinfo {author} {\bibfnamefont {H.}~\bibnamefont {Schomerus}}, \ and\
  \bibinfo {author} {\bibfnamefont {C.W.J.}\ \bibnamefont {Beenakker}},\
  }\bibfield  {title} {\enquote {\bibinfo {title} {{A zero-voltage conductance
  peak from weak antilocalization in a Majorana nanowire}},}\ }\href
  {http://stacks.iop.org/1367-2630/14/i=12/a=125011} {\bibfield  {journal}
  {\bibinfo  {journal} {New Journal of Physics}\ }\textbf {\bibinfo {volume}
  {14}},\ \bibinfo {pages} {125011} (\bibinfo {year} {2012})}\BibitemShut
  {NoStop}%
\bibitem [{Note1()}]{Note1}%
  \BibitemOpen
  \bibinfo {note} {See Supplemental Material for data from additional samples
  and additional micrographs of the main sample}\BibitemShut {NoStop}%
\bibitem [{\citenamefont {Analytis}\ \emph {et~al.}(2010)\citenamefont
  {Analytis}, \citenamefont {McDonald}, \citenamefont {Riggs}, \citenamefont
  {Chu}, \citenamefont {Boebinger},\ and\ \citenamefont
  {Fisher}}]{Nat.Phys.6.960}%
  \BibitemOpen
  \bibfield  {author} {\bibinfo {author} {\bibfnamefont {James~G.}\
  \bibnamefont {Analytis}}, \bibinfo {author} {\bibfnamefont {Ross~D.}\
  \bibnamefont {McDonald}}, \bibinfo {author} {\bibfnamefont {Scott~C.}\
  \bibnamefont {Riggs}}, \bibinfo {author} {\bibfnamefont {Jiun-Haw}\
  \bibnamefont {Chu}}, \bibinfo {author} {\bibfnamefont {G.~S.}\ \bibnamefont
  {Boebinger}}, \ and\ \bibinfo {author} {\bibfnamefont {Ian~R.}\ \bibnamefont
  {Fisher}},\ }\bibfield  {title} {\enquote {\bibinfo {title} {{Two-dimensional
  surface state in the quantum limit of a topological insulator}},}\ }\href
  {\doibase 10.1038/nphys1861} {\bibfield  {journal} {\bibinfo  {journal} {Nat.
  Phys.}\ }\textbf {\bibinfo {volume} {6}},\ \bibinfo {pages} {960} (\bibinfo
  {year} {2010})}\BibitemShut {NoStop}%
\bibitem [{\citenamefont {Blonder}\ \emph {et~al.}(1982)\citenamefont
  {Blonder}, \citenamefont {Tinkham},\ and\ \citenamefont
  {Klapwijk}}]{PhysRevB.25.4515}%
  \BibitemOpen
  \bibfield  {author} {\bibinfo {author} {\bibfnamefont {G.~E.}\ \bibnamefont
  {Blonder}}, \bibinfo {author} {\bibfnamefont {M.}~\bibnamefont {Tinkham}}, \
  and\ \bibinfo {author} {\bibfnamefont {T.~M.}\ \bibnamefont {Klapwijk}},\
  }\bibfield  {title} {\enquote {\bibinfo {title} {{Transition from metallic to
  tunneling regimes in superconducting microconstrictions: Excess current,
  charge imbalance, and supercurrent conversion}},}\ }\href {\doibase
  10.1103/PhysRevB.25.4515} {\bibfield  {journal} {\bibinfo  {journal} {Phys.
  Rev. B}\ }\textbf {\bibinfo {volume} {25}},\ \bibinfo {pages} {4515--4532}
  (\bibinfo {year} {1982})}\BibitemShut {NoStop}%
\bibitem [{\citenamefont {Yang}\ \emph {et~al.}(2012)\citenamefont {Yang},
  \citenamefont {Ding}, \citenamefont {Qu}, \citenamefont {Shen}, \citenamefont
  {Chen}, \citenamefont {Wei}, \citenamefont {Ji}, \citenamefont {Liu},
  \citenamefont {Fan}, \citenamefont {Yang}, \citenamefont {Xiang},\ and\
  \citenamefont {Lu}}]{PhysRevB.85.104508}%
  \BibitemOpen
  \bibfield  {author} {\bibinfo {author} {\bibfnamefont {Fan}\ \bibnamefont
  {Yang}}, \bibinfo {author} {\bibfnamefont {Yue}\ \bibnamefont {Ding}},
  \bibinfo {author} {\bibfnamefont {Fanming}\ \bibnamefont {Qu}}, \bibinfo
  {author} {\bibfnamefont {Jie}\ \bibnamefont {Shen}}, \bibinfo {author}
  {\bibfnamefont {Jun}\ \bibnamefont {Chen}}, \bibinfo {author} {\bibfnamefont
  {Zhongchao}\ \bibnamefont {Wei}}, \bibinfo {author} {\bibfnamefont
  {Zhongqing}\ \bibnamefont {Ji}}, \bibinfo {author} {\bibfnamefont
  {Guangtong}\ \bibnamefont {Liu}}, \bibinfo {author} {\bibfnamefont {Jie}\
  \bibnamefont {Fan}}, \bibinfo {author} {\bibfnamefont {Changli}\ \bibnamefont
  {Yang}}, \bibinfo {author} {\bibfnamefont {Tao}\ \bibnamefont {Xiang}}, \
  and\ \bibinfo {author} {\bibfnamefont {Li}~\bibnamefont {Lu}},\ }\bibfield
  {title} {\enquote {\bibinfo {title} {{Proximity effect at superconducting
  Sn-Bi${}_{2}$Se${}_{3}$ interface}},}\ }\href {\doibase
  10.1103/PhysRevB.85.104508} {\bibfield  {journal} {\bibinfo  {journal} {Phys.
  Rev. B}\ }\textbf {\bibinfo {volume} {85}},\ \bibinfo {pages} {104508}
  (\bibinfo {year} {2012})}\BibitemShut {NoStop}%
\bibitem [{\citenamefont {Koren}\ and\ \citenamefont
  {Kirzhner}(2012)}]{PhysRevB.86.144508}%
  \BibitemOpen
  \bibfield  {author} {\bibinfo {author} {\bibfnamefont {G.}~\bibnamefont
  {Koren}}\ and\ \bibinfo {author} {\bibfnamefont {T.}~\bibnamefont
  {Kirzhner}},\ }\bibfield  {title} {\enquote {\bibinfo {title} {{Zero-energy
  bound states in tunneling conductance spectra at the interface of an $s$-wave
  superconductor and a topological insulator in NbN/Bi${}_{2}$Se${}_{3}$/Au
  thin-film junctions}},}\ }\href {\doibase 10.1103/PhysRevB.86.144508}
  {\bibfield  {journal} {\bibinfo  {journal} {Phys. Rev. B}\ }\textbf {\bibinfo
  {volume} {86}},\ \bibinfo {pages} {144508} (\bibinfo {year}
  {2012})}\BibitemShut {NoStop}%
\bibitem [{\citenamefont {Artemenko}\ \emph {et~al.}(1979)\citenamefont
  {Artemenko}, \citenamefont {Volkov},\ and\ \citenamefont
  {Zaitsev}}]{Artemenko1979771}%
  \BibitemOpen
  \bibfield  {author} {\bibinfo {author} {\bibfnamefont {S.N.}\ \bibnamefont
  {Artemenko}}, \bibinfo {author} {\bibfnamefont {A.F.}\ \bibnamefont
  {Volkov}}, \ and\ \bibinfo {author} {\bibfnamefont {A.V.}\ \bibnamefont
  {Zaitsev}},\ }\bibfield  {title} {\enquote {\bibinfo {title} {On the excess
  current in microbridges s-c-s and s-c-n},}\ }\href {\doibase
  http://dx.doi.org/10.1016/0038-1098(79)90044-9} {\bibfield  {journal}
  {\bibinfo  {journal} {Solid State Communications}\ }\textbf {\bibinfo
  {volume} {30}},\ \bibinfo {pages} {771 -- 773} (\bibinfo {year}
  {1979})}\BibitemShut {NoStop}%
\bibitem [{\citenamefont {Nazarov}\ and\ \citenamefont
  {Stoof}(1996)}]{PhysRevLett.76.823}%
  \BibitemOpen
  \bibfield  {author} {\bibinfo {author} {\bibfnamefont {Yuli~V.}\ \bibnamefont
  {Nazarov}}\ and\ \bibinfo {author} {\bibfnamefont {T.~H.}\ \bibnamefont
  {Stoof}},\ }\bibfield  {title} {\enquote {\bibinfo {title} {{Diffusive
  Conductors as Andreev Interferometers}},}\ }\href {\doibase
  10.1103/PhysRevLett.76.823} {\bibfield  {journal} {\bibinfo  {journal} {Phys.
  Rev. Lett.}\ }\textbf {\bibinfo {volume} {76}},\ \bibinfo {pages} {823--826}
  (\bibinfo {year} {1996})}\BibitemShut {NoStop}%
\bibitem [{\citenamefont {Golubov}\ \emph {et~al.}(1997)\citenamefont
  {Golubov}, \citenamefont {Wilhelm},\ and\ \citenamefont
  {Zaikin}}]{PhysRevB.55.1123}%
  \BibitemOpen
  \bibfield  {author} {\bibinfo {author} {\bibfnamefont {A.~A.}\ \bibnamefont
  {Golubov}}, \bibinfo {author} {\bibfnamefont {F.~K.}\ \bibnamefont
  {Wilhelm}}, \ and\ \bibinfo {author} {\bibfnamefont {A.~D.}\ \bibnamefont
  {Zaikin}},\ }\bibfield  {title} {\enquote {\bibinfo {title} {{Coherent charge
  transport in metallic proximity structures}},}\ }\href {\doibase
  10.1103/PhysRevB.55.1123} {\bibfield  {journal} {\bibinfo  {journal} {Phys.
  Rev. B}\ }\textbf {\bibinfo {volume} {55}},\ \bibinfo {pages} {1123--1137}
  (\bibinfo {year} {1997})}\BibitemShut {NoStop}%
\bibitem [{\citenamefont {Hsieh}\ and\ \citenamefont
  {Fu}(2012)}]{PhysRevLett.108.107005}%
  \BibitemOpen
  \bibfield  {author} {\bibinfo {author} {\bibfnamefont {Timothy~H.}\
  \bibnamefont {Hsieh}}\ and\ \bibinfo {author} {\bibfnamefont {Liang}\
  \bibnamefont {Fu}},\ }\bibfield  {title} {\enquote {\bibinfo {title}
  {{Majorana Fermions and Exotic Surface Andreev Bound States in Topological
  Superconductors: Application to Cu$_x$Bi$_2$Se$_3$}},}\ }\href {\doibase
  10.1103/PhysRevLett.108.107005} {\bibfield  {journal} {\bibinfo  {journal}
  {Phys. Rev. Lett.}\ }\textbf {\bibinfo {volume} {108}},\ \bibinfo {pages}
  {107005} (\bibinfo {year} {2012})}\BibitemShut {NoStop}%
\bibitem [{\citenamefont {Sasaki}\ \emph {et~al.}(2011)\citenamefont {Sasaki},
  \citenamefont {Kriener}, \citenamefont {Segawa}, \citenamefont {Yada},
  \citenamefont {Tanaka}, \citenamefont {Sato},\ and\ \citenamefont
  {Ando}}]{PhysRevLett.107.217001}%
  \BibitemOpen
  \bibfield  {author} {\bibinfo {author} {\bibfnamefont {Satoshi}\ \bibnamefont
  {Sasaki}}, \bibinfo {author} {\bibfnamefont {M.}~\bibnamefont {Kriener}},
  \bibinfo {author} {\bibfnamefont {Kouji}\ \bibnamefont {Segawa}}, \bibinfo
  {author} {\bibfnamefont {Keiji}\ \bibnamefont {Yada}}, \bibinfo {author}
  {\bibfnamefont {Yukio}\ \bibnamefont {Tanaka}}, \bibinfo {author}
  {\bibfnamefont {Masatoshi}\ \bibnamefont {Sato}}, \ and\ \bibinfo {author}
  {\bibfnamefont {Yoichi}\ \bibnamefont {Ando}},\ }\bibfield  {title} {\enquote
  {\bibinfo {title} {{Topological Superconductivity in Cu$_x$Bi$_2$Se$_3$}},}\
  }\href {\doibase 10.1103/PhysRevLett.107.217001} {\bibfield  {journal}
  {\bibinfo  {journal} {Phys. Rev. Lett.}\ }\textbf {\bibinfo {volume} {107}},\
  \bibinfo {pages} {217001} (\bibinfo {year} {2011})}\BibitemShut {NoStop}%
\bibitem [{\citenamefont {Checkelsky}\ \emph {et~al.}(2011)\citenamefont
  {Checkelsky}, \citenamefont {Hor}, \citenamefont {Cava},\ and\ \citenamefont
  {Ong}}]{PhysRevLett.106.196801}%
  \BibitemOpen
  \bibfield  {author} {\bibinfo {author} {\bibfnamefont {J.~G.}\ \bibnamefont
  {Checkelsky}}, \bibinfo {author} {\bibfnamefont {Y.~S.}\ \bibnamefont {Hor}},
  \bibinfo {author} {\bibfnamefont {R.~J.}\ \bibnamefont {Cava}}, \ and\
  \bibinfo {author} {\bibfnamefont {N.~P.}\ \bibnamefont {Ong}},\ }\bibfield
  {title} {\enquote {\bibinfo {title} {{Bulk Band Gap and Surface State
  Conduction Observed in Voltage-Tuned Crystals of the Topological Insulator
  ${\mathrm{Bi}}_{2}{\mathrm{Se}}_{3}$}},}\ }\href {\doibase
  10.1103/PhysRevLett.106.196801} {\bibfield  {journal} {\bibinfo  {journal}
  {Phys. Rev. Lett.}\ }\textbf {\bibinfo {volume} {106}},\ \bibinfo {pages}
  {196801} (\bibinfo {year} {2011})}\BibitemShut {NoStop}%
\bibitem [{\citenamefont {Kim}\ \emph {et~al.}({2013})\citenamefont {Kim},
  \citenamefont {Syers}, \citenamefont {Butch}, \citenamefont {Paglione},\ and\
  \citenamefont {Fuhrer}}]{NatComm.4.2040}%
  \BibitemOpen
  \bibfield  {author} {\bibinfo {author} {\bibfnamefont {Dohun}\ \bibnamefont
  {Kim}}, \bibinfo {author} {\bibfnamefont {Paul}\ \bibnamefont {Syers}},
  \bibinfo {author} {\bibfnamefont {Nicholas~P.}\ \bibnamefont {Butch}},
  \bibinfo {author} {\bibfnamefont {Johnpierre}\ \bibnamefont {Paglione}}, \
  and\ \bibinfo {author} {\bibfnamefont {Michael~S.}\ \bibnamefont {Fuhrer}},\
  }\bibfield  {title} {\enquote {\bibinfo {title} {{ Coherent topological
  transport on the surface of Bi$_2$Se$_3$}},}\ }\href {\doibase
  {10.1038/ncomms3040}} {\bibfield  {journal} {\bibinfo  {journal} {{Nat.
  Comm.}}\ }\textbf {\bibinfo {volume} {{4}}},\ \bibinfo {pages} {{2040}}
  (\bibinfo {year} {{2013}})}\BibitemShut {NoStop}%
\bibitem [{\citenamefont {Bolotin}\ \emph {et~al.}({2008})\citenamefont
  {Bolotin}, \citenamefont {Sikes}, \citenamefont {Jiang}, \citenamefont
  {Klima}, \citenamefont {Fudenberg}, \citenamefont {Hone}, \citenamefont
  {Kim},\ and\ \citenamefont {Stormer}}]{Bolotin2008351}%
  \BibitemOpen
  \bibfield  {author} {\bibinfo {author} {\bibfnamefont {K.I.}\ \bibnamefont
  {Bolotin}}, \bibinfo {author} {\bibfnamefont {K.J.}\ \bibnamefont {Sikes}},
  \bibinfo {author} {\bibfnamefont {Z.}~\bibnamefont {Jiang}}, \bibinfo
  {author} {\bibfnamefont {M.}~\bibnamefont {Klima}}, \bibinfo {author}
  {\bibfnamefont {G.}~\bibnamefont {Fudenberg}}, \bibinfo {author}
  {\bibfnamefont {J.}~\bibnamefont {Hone}}, \bibinfo {author} {\bibfnamefont
  {P.}~\bibnamefont {Kim}}, \ and\ \bibinfo {author} {\bibfnamefont {H.L.}\
  \bibnamefont {Stormer}},\ }\bibfield  {title} {\enquote {\bibinfo {title}
  {{Ultrahigh electron mobility in suspended graphene}},}\ }\href {\doibase
  {http://dx.doi.org/10.1016/j.ssc.2008.02.024}} {\bibfield  {journal}
  {\bibinfo  {journal} {{Solid State Commun.}}\ }\textbf {\bibinfo {volume}
  {{146}}},\ \bibinfo {pages} {{351 -- 355}} (\bibinfo {year}
  {{2008}})}\BibitemShut {NoStop}%
\bibitem [{\citenamefont {Bianchi}\ \emph {et~al.}(2010)\citenamefont
  {Bianchi}, \citenamefont {Guan}, \citenamefont {Bao}, \citenamefont {Mi},
  \citenamefont {Iversen}, \citenamefont {King},\ and\ \citenamefont
  {Hofmann}}]{NatComm.1.128}%
  \BibitemOpen
  \bibfield  {author} {\bibinfo {author} {\bibfnamefont {Marco}\ \bibnamefont
  {Bianchi}}, \bibinfo {author} {\bibfnamefont {Dandan}\ \bibnamefont {Guan}},
  \bibinfo {author} {\bibfnamefont {Shining}\ \bibnamefont {Bao}}, \bibinfo
  {author} {\bibfnamefont {Jianli}\ \bibnamefont {Mi}}, \bibinfo {author}
  {\bibfnamefont {Bo~Brummerstedt}\ \bibnamefont {Iversen}}, \bibinfo {author}
  {\bibfnamefont {Philip~D.C.}\ \bibnamefont {King}}, \ and\ \bibinfo {author}
  {\bibfnamefont {Philip}\ \bibnamefont {Hofmann}},\ }\bibfield  {title}
  {\enquote {\bibinfo {title} {{Coexistence of the topological state and a
  two-dimensional electron gas on the surface of Bi$_2$Se$_3$}},}\ }\href
  {\doibase 10.1038/ncomms1131} {\bibfield  {journal} {\bibinfo  {journal}
  {Nat. Comm.}\ }\textbf {\bibinfo {volume} {1}},\ \bibinfo {pages} {128}
  (\bibinfo {year} {2010})}\BibitemShut {NoStop}%
\bibitem [{\citenamefont {Bansal}\ \emph {et~al.}(2012)\citenamefont {Bansal},
  \citenamefont {Kim}, \citenamefont {Brahlek}, \citenamefont {Edrey},\ and\
  \citenamefont {Oh}}]{PhysRevLett.109.116804}%
  \BibitemOpen
  \bibfield  {author} {\bibinfo {author} {\bibfnamefont {Namrata}\ \bibnamefont
  {Bansal}}, \bibinfo {author} {\bibfnamefont {Yong~Seung}\ \bibnamefont
  {Kim}}, \bibinfo {author} {\bibfnamefont {Matthew}\ \bibnamefont {Brahlek}},
  \bibinfo {author} {\bibfnamefont {Eliav}\ \bibnamefont {Edrey}}, \ and\
  \bibinfo {author} {\bibfnamefont {Seongshik}\ \bibnamefont {Oh}},\ }\bibfield
   {title} {\enquote {\bibinfo {title} {{Thickness-Independent Transport
  Channels in Topological Insulator Bi$_2$Se$_3$ Thin Films}},}\ }\href
  {\doibase 10.1103/PhysRevLett.109.116804} {\bibfield  {journal} {\bibinfo
  {journal} {Phys. Rev. Lett.}\ }\textbf {\bibinfo {volume} {109}},\ \bibinfo
  {pages} {116804} (\bibinfo {year} {2012})}\BibitemShut {NoStop}%
\bibitem [{\citenamefont {Cao}\ \emph {et~al.}(2005)\citenamefont {Cao},
  \citenamefont {Wang},\ and\ \citenamefont {Dai}}]{NatMater.4.745}%
  \BibitemOpen
  \bibfield  {author} {\bibinfo {author} {\bibfnamefont {Jien}\ \bibnamefont
  {Cao}}, \bibinfo {author} {\bibfnamefont {Qian}\ \bibnamefont {Wang}}, \ and\
  \bibinfo {author} {\bibfnamefont {Hongjie}\ \bibnamefont {Dai}},\ }\bibfield
  {title} {\enquote {\bibinfo {title} {{Electron transport in very clean,
  as-grown suspended carbon nanotubes}},}\ }\href {\doibase 10.1038/nmat1478}
  {\bibfield  {journal} {\bibinfo  {journal} {Nat. Mater.}\ }\textbf {\bibinfo
  {volume} {4}},\ \bibinfo {pages} {745} (\bibinfo {year} {2005})}\BibitemShut
  {NoStop}%
\bibitem [{\citenamefont {Giazotto}\ \emph {et~al.}(2001)\citenamefont
  {Giazotto}, \citenamefont {Pingue}, \citenamefont {Beltram}, \citenamefont
  {Lazzarino}, \citenamefont {Orani}, \citenamefont {Rubini},\ and\
  \citenamefont {Franciosi}}]{PhysRevLett.87.216808}%
  \BibitemOpen
  \bibfield  {author} {\bibinfo {author} {\bibfnamefont {Francesco}\
  \bibnamefont {Giazotto}}, \bibinfo {author} {\bibfnamefont {Pasqualantonio}\
  \bibnamefont {Pingue}}, \bibinfo {author} {\bibfnamefont {Fabio}\
  \bibnamefont {Beltram}}, \bibinfo {author} {\bibfnamefont {Marco}\
  \bibnamefont {Lazzarino}}, \bibinfo {author} {\bibfnamefont {Daniela}\
  \bibnamefont {Orani}}, \bibinfo {author} {\bibfnamefont {Silvia}\
  \bibnamefont {Rubini}}, \ and\ \bibinfo {author} {\bibfnamefont {Alfonso}\
  \bibnamefont {Franciosi}},\ }\bibfield  {title} {\enquote {\bibinfo {title}
  {{Resonant Transport in Nb/GaAs/AlGaAs Heterostructures: Realization of the
  de GennesÐSaint-James Model}},}\ }\href {\doibase
  10.1103/PhysRevLett.87.216808} {\bibfield  {journal} {\bibinfo  {journal}
  {Phys. Rev. Lett.}\ }\textbf {\bibinfo {volume} {87}},\ \bibinfo {pages}
  {216808} (\bibinfo {year} {2001})}\BibitemShut {NoStop}%
\bibitem [{\citenamefont {Levi}\ \emph {et~al.}({1999})\citenamefont {Levi},
  \citenamefont {Millo}, \citenamefont {Rizzo}, \citenamefont {Prober},\ and\
  \citenamefont {Motowidlo}}]{ApplSurfSci.144.575}%
  \BibitemOpen
  \bibfield  {author} {\bibinfo {author} {\bibfnamefont {Y}~\bibnamefont
  {Levi}}, \bibinfo {author} {\bibfnamefont {O}~\bibnamefont {Millo}}, \bibinfo
  {author} {\bibfnamefont {N.D}\ \bibnamefont {Rizzo}}, \bibinfo {author}
  {\bibfnamefont {D.E}\ \bibnamefont {Prober}}, \ and\ \bibinfo {author}
  {\bibfnamefont {L.R}\ \bibnamefont {Motowidlo}},\ }\bibfield  {title}
  {\enquote {\bibinfo {title} {{Tunneling spectroscopy of bound and resonant
  states in superconducting proximity structures}},}\ }\href {\doibase
  {http://dx.doi.org/10.1016/S0169-4332(98)00868-X}} {\bibfield  {journal}
  {\bibinfo  {journal} {{Appl. Surf. Sc.}}\ }\textbf {\bibinfo {volume}
  {{144-145}}},\ \bibinfo {pages} {{575 -- 579}} (\bibinfo {year}
  {{1999}})}\BibitemShut {NoStop}%
\bibitem [{\citenamefont {Bird}\ \emph {et~al.}(1995)\citenamefont {Bird},
  \citenamefont {Ishibashi}, \citenamefont {Ferry}, \citenamefont {Ochiai},
  \citenamefont {Aoyagi},\ and\ \citenamefont {Sugano}}]{PhysRevB.51.18037}%
  \BibitemOpen
  \bibfield  {author} {\bibinfo {author} {\bibfnamefont {J.~P.}\ \bibnamefont
  {Bird}}, \bibinfo {author} {\bibfnamefont {K.}~\bibnamefont {Ishibashi}},
  \bibinfo {author} {\bibfnamefont {D.~K.}\ \bibnamefont {Ferry}}, \bibinfo
  {author} {\bibfnamefont {Y.}~\bibnamefont {Ochiai}}, \bibinfo {author}
  {\bibfnamefont {Y.}~\bibnamefont {Aoyagi}}, \ and\ \bibinfo {author}
  {\bibfnamefont {T.}~\bibnamefont {Sugano}},\ }\bibfield  {title} {\enquote
  {\bibinfo {title} {{Phase breaking in ballistic quantum dots: Transition from
  two- to zero-dimensional behavior}},}\ }\href {\doibase
  10.1103/PhysRevB.51.18037} {\bibfield  {journal} {\bibinfo  {journal} {Phys.
  Rev. B}\ }\textbf {\bibinfo {volume} {51}},\ \bibinfo {pages} {18037--18040}
  (\bibinfo {year} {1995})}\BibitemShut {NoStop}%
\bibitem [{\citenamefont {Huibers}\ \emph {et~al.}(1998)\citenamefont
  {Huibers}, \citenamefont {Switkes}, \citenamefont {Marcus}, \citenamefont
  {Campman},\ and\ \citenamefont {Gossard}}]{PhysRevLett.81.200}%
  \BibitemOpen
  \bibfield  {author} {\bibinfo {author} {\bibfnamefont {A.~G.}\ \bibnamefont
  {Huibers}}, \bibinfo {author} {\bibfnamefont {M.}~\bibnamefont {Switkes}},
  \bibinfo {author} {\bibfnamefont {C.~M.}\ \bibnamefont {Marcus}}, \bibinfo
  {author} {\bibfnamefont {K.}~\bibnamefont {Campman}}, \ and\ \bibinfo
  {author} {\bibfnamefont {A.~C.}\ \bibnamefont {Gossard}},\ }\bibfield
  {title} {\enquote {\bibinfo {title} {{Dephasing in Open Quantum Dots}},}\
  }\href {\doibase 10.1103/PhysRevLett.81.200} {\bibfield  {journal} {\bibinfo
  {journal} {Phys. Rev. Lett.}\ }\textbf {\bibinfo {volume} {81}},\ \bibinfo
  {pages} {200--203} (\bibinfo {year} {1998})}\BibitemShut {NoStop}%
\bibitem [{\citenamefont {Hansen}\ \emph {et~al.}(2001)\citenamefont {Hansen},
  \citenamefont {Kristensen}, \citenamefont {Pedersen}, \citenamefont
  {S\o{}rensen},\ and\ \citenamefont {Lindelof}}]{PhysRevB.64.045327}%
  \BibitemOpen
  \bibfield  {author} {\bibinfo {author} {\bibfnamefont {A.~E.}\ \bibnamefont
  {Hansen}}, \bibinfo {author} {\bibfnamefont {A.}~\bibnamefont {Kristensen}},
  \bibinfo {author} {\bibfnamefont {S.}~\bibnamefont {Pedersen}}, \bibinfo
  {author} {\bibfnamefont {C.~B.}\ \bibnamefont {S\o{}rensen}}, \ and\ \bibinfo
  {author} {\bibfnamefont {P.~E.}\ \bibnamefont {Lindelof}},\ }\bibfield
  {title} {\enquote {\bibinfo {title} {{Mesoscopic decoherence in Aharonov-Bohm
  rings}},}\ }\href {\doibase 10.1103/PhysRevB.64.045327} {\bibfield  {journal}
  {\bibinfo  {journal} {Phys. Rev. B}\ }\textbf {\bibinfo {volume} {64}},\
  \bibinfo {pages} {045327} (\bibinfo {year} {2001})}\BibitemShut {NoStop}%
\bibitem [{\citenamefont {Lin}\ \emph {et~al.}(2010)\citenamefont {Lin},
  \citenamefont {Lin}, \citenamefont {Chi}, \citenamefont {Chen}, \citenamefont
  {Ueda},\ and\ \citenamefont {Komiyama}}]{PhysRevB.81.035312}%
  \BibitemOpen
  \bibfield  {author} {\bibinfo {author} {\bibfnamefont {Kuan-Ting}\
  \bibnamefont {Lin}}, \bibinfo {author} {\bibfnamefont {Yiping}\ \bibnamefont
  {Lin}}, \bibinfo {author} {\bibfnamefont {C.~C.}\ \bibnamefont {Chi}},
  \bibinfo {author} {\bibfnamefont {J.~C.}\ \bibnamefont {Chen}}, \bibinfo
  {author} {\bibfnamefont {T.}~\bibnamefont {Ueda}}, \ and\ \bibinfo {author}
  {\bibfnamefont {S.}~\bibnamefont {Komiyama}},\ }\bibfield  {title} {\enquote
  {\bibinfo {title} {{Temperature- and current-dependent dephasing in an
  Aharonov-Bohm ring}},}\ }\href {\doibase 10.1103/PhysRevB.81.035312}
  {\bibfield  {journal} {\bibinfo  {journal} {Phys. Rev. B}\ }\textbf {\bibinfo
  {volume} {81}},\ \bibinfo {pages} {035312} (\bibinfo {year}
  {2010})}\BibitemShut {NoStop}%
\bibitem [{\citenamefont {Bao}\ \emph {et~al.}(2012)\citenamefont {Bao},
  \citenamefont {He}, \citenamefont {Meyer}, \citenamefont {Kou}, \citenamefont
  {Zhang}, \citenamefont {Chen}, \citenamefont {Fedorov}, \citenamefont {Zou},
  \citenamefont {Riedemann}, \citenamefont {Lograsso}, \citenamefont {Wang},
  \citenamefont {Tuttle},\ and\ \citenamefont {Xiu}}]{SciRep.2.726}%
  \BibitemOpen
  \bibfield  {author} {\bibinfo {author} {\bibfnamefont {Lihong}\ \bibnamefont
  {Bao}}, \bibinfo {author} {\bibfnamefont {Liang}\ \bibnamefont {He}},
  \bibinfo {author} {\bibfnamefont {Nicholas}\ \bibnamefont {Meyer}}, \bibinfo
  {author} {\bibfnamefont {Xufeng}\ \bibnamefont {Kou}}, \bibinfo {author}
  {\bibfnamefont {Peng}\ \bibnamefont {Zhang}}, \bibinfo {author}
  {\bibfnamefont {Zhi-gang}\ \bibnamefont {Chen}}, \bibinfo {author}
  {\bibfnamefont {Alexei~V.}\ \bibnamefont {Fedorov}}, \bibinfo {author}
  {\bibfnamefont {Jin}\ \bibnamefont {Zou}}, \bibinfo {author} {\bibfnamefont
  {Trevor~M.}\ \bibnamefont {Riedemann}}, \bibinfo {author} {\bibfnamefont
  {Thomas~A.}\ \bibnamefont {Lograsso}}, \bibinfo {author} {\bibfnamefont
  {Kang~L.}\ \bibnamefont {Wang}}, \bibinfo {author} {\bibfnamefont {Gary}\
  \bibnamefont {Tuttle}}, \ and\ \bibinfo {author} {\bibfnamefont {Faxian}\
  \bibnamefont {Xiu}},\ }\bibfield  {title} {\enquote {\bibinfo {title} {{Weak
  Anti-localization and Quantum Oscillations of Surface States in Topological
  Insulator Bi$_2$Se$_2$Te}},}\ }\href {\doibase 10.1038/srep00726} {\bibfield
  {journal} {\bibinfo  {journal} {Sci. Rep.}\ }\textbf {\bibinfo {volume}
  {2}},\ \bibinfo {pages} {726} (\bibinfo {year} {2012})}\BibitemShut {NoStop}%
\bibitem [{\citenamefont {Kim}\ \emph {et~al.}({2012})\citenamefont {Kim},
  \citenamefont {Cho}, \citenamefont {Butch}, \citenamefont {Syers},
  \citenamefont {Kirshenbaum}, \citenamefont {Adam}, \citenamefont {Paglione},\
  and\ \citenamefont {Fuhrer}}]{NatPhys.8.460}%
  \BibitemOpen
  \bibfield  {author} {\bibinfo {author} {\bibfnamefont {Dohun}\ \bibnamefont
  {Kim}}, \bibinfo {author} {\bibfnamefont {Sungjae}\ \bibnamefont {Cho}},
  \bibinfo {author} {\bibfnamefont {Nicholas~P.}\ \bibnamefont {Butch}},
  \bibinfo {author} {\bibfnamefont {Paul}\ \bibnamefont {Syers}}, \bibinfo
  {author} {\bibfnamefont {Kevin}\ \bibnamefont {Kirshenbaum}}, \bibinfo
  {author} {\bibfnamefont {Shaffique}\ \bibnamefont {Adam}}, \bibinfo {author}
  {\bibfnamefont {Johnpierre}\ \bibnamefont {Paglione}}, \ and\ \bibinfo
  {author} {\bibfnamefont {Michael~S.}\ \bibnamefont {Fuhrer}},\ }\bibfield
  {title} {\enquote {\bibinfo {title} {{Surface conduction of topological Dirac
  electrons in bulk insulating Bi$_2$Se$_3$}},}\ }\href {\doibase
  {10.1038/nphys2286}} {\bibfield  {journal} {\bibinfo  {journal} {{Nat.
  Phys.}}\ }\textbf {\bibinfo {volume} {{8}}},\ \bibinfo {pages} {{460}}
  (\bibinfo {year} {{2012}})}\BibitemShut {NoStop}%
\bibitem [{\citenamefont {Shytov}\ \emph {et~al.}(2008)\citenamefont {Shytov},
  \citenamefont {Rudner},\ and\ \citenamefont
  {Levitov}}]{PhysRevLett.101.156804}%
  \BibitemOpen
  \bibfield  {author} {\bibinfo {author} {\bibfnamefont {Andrei~V.}\
  \bibnamefont {Shytov}}, \bibinfo {author} {\bibfnamefont {Mark~S.}\
  \bibnamefont {Rudner}}, \ and\ \bibinfo {author} {\bibfnamefont {Leonid~S.}\
  \bibnamefont {Levitov}},\ }\bibfield  {title} {\enquote {\bibinfo {title}
  {{Klein Backscattering and Fabry-P\'{e}rot Interference in Graphene
  Heterojunctions}},}\ }\href {\doibase 10.1103/PhysRevLett.101.156804}
  {\bibfield  {journal} {\bibinfo  {journal} {Phys. Rev. Lett.}\ }\textbf
  {\bibinfo {volume} {101}},\ \bibinfo {pages} {156804} (\bibinfo {year}
  {2008})}\BibitemShut {NoStop}%
\bibitem [{\citenamefont {van Wees}\ \emph {et~al.}(1992)\citenamefont {van
  Wees}, \citenamefont {de~Vries}, \citenamefont {Magn\'ee},\ and\
  \citenamefont {Klapwijk}}]{PhysRevLett.69.510}%
  \BibitemOpen
  \bibfield  {author} {\bibinfo {author} {\bibfnamefont {B.~J.}\ \bibnamefont
  {van Wees}}, \bibinfo {author} {\bibfnamefont {P.}~\bibnamefont {de~Vries}},
  \bibinfo {author} {\bibfnamefont {P.}~\bibnamefont {Magn\'ee}}, \ and\
  \bibinfo {author} {\bibfnamefont {T.~M.}\ \bibnamefont {Klapwijk}},\
  }\bibfield  {title} {\enquote {\bibinfo {title} {{Excess conductance of
  superconductor-semiconductor interfaces due to phase conjugation between
  electrons and holes}},}\ }\href {\doibase 10.1103/PhysRevLett.69.510}
  {\bibfield  {journal} {\bibinfo  {journal} {Phys. Rev. Lett.}\ }\textbf
  {\bibinfo {volume} {69}},\ \bibinfo {pages} {510--513} (\bibinfo {year}
  {1992})}\BibitemShut {NoStop}%
\bibitem [{\citenamefont {Schechter}\ \emph {et~al.}(2001)\citenamefont
  {Schechter}, \citenamefont {Imry},\ and\ \citenamefont
  {Levinson}}]{PhysRevB.64.224513}%
  \BibitemOpen
  \bibfield  {author} {\bibinfo {author} {\bibfnamefont {M.}~\bibnamefont
  {Schechter}}, \bibinfo {author} {\bibfnamefont {Y.}~\bibnamefont {Imry}}, \
  and\ \bibinfo {author} {\bibfnamefont {Y.}~\bibnamefont {Levinson}},\
  }\bibfield  {title} {\enquote {\bibinfo {title} {{Reflectionless tunneling in
  ballistic normal$-$metal$-$superconductor junctions}},}\ }\href {\doibase
  10.1103/PhysRevB.64.224513} {\bibfield  {journal} {\bibinfo  {journal} {Phys.
  Rev. B}\ }\textbf {\bibinfo {volume} {64}},\ \bibinfo {pages} {224513}
  (\bibinfo {year} {2001})}\BibitemShut {NoStop}%
\bibitem [{\citenamefont {Lee}\ \emph {et~al.}(2012)\citenamefont {Lee},
  \citenamefont {Jiang}, \citenamefont {Aguado}, \citenamefont {Katsaros},
  \citenamefont {Lieber},\ and\ \citenamefont
  {De~Franceschi}}]{PhysRevLett.109.186802}%
  \BibitemOpen
  \bibfield  {author} {\bibinfo {author} {\bibfnamefont {Eduardo J.~H.}\
  \bibnamefont {Lee}}, \bibinfo {author} {\bibfnamefont {Xiaocheng}\
  \bibnamefont {Jiang}}, \bibinfo {author} {\bibfnamefont {Ram\'on}\
  \bibnamefont {Aguado}}, \bibinfo {author} {\bibfnamefont {Georgios}\
  \bibnamefont {Katsaros}}, \bibinfo {author} {\bibfnamefont {Charles~M.}\
  \bibnamefont {Lieber}}, \ and\ \bibinfo {author} {\bibfnamefont {Silvano}\
  \bibnamefont {De~Franceschi}},\ }\bibfield  {title} {\enquote {\bibinfo
  {title} {{Zero-Bias Anomaly in a Nanowire Quantum Dot Coupled to
  Superconductors}},}\ }\href {\doibase 10.1103/PhysRevLett.109.186802}
  {\bibfield  {journal} {\bibinfo  {journal} {Phys. Rev. Lett.}\ }\textbf
  {\bibinfo {volume} {109}},\ \bibinfo {pages} {186802} (\bibinfo {year}
  {2012})}\BibitemShut {NoStop}%
\bibitem [{\citenamefont {Lee}\ \emph {et~al.}(2014)\citenamefont {Lee},
  \citenamefont {Jiang}, \citenamefont {Houzet}, \citenamefont {Aguardo},
  \citenamefont {Lieber},\ and\ \citenamefont
  {De~Franceschi}}]{NatureNanotechnology.9.79}%
  \BibitemOpen
  \bibfield  {author} {\bibinfo {author} {\bibfnamefont {Eduardo~J.H.}\
  \bibnamefont {Lee}}, \bibinfo {author} {\bibfnamefont {Xiaocheng}\
  \bibnamefont {Jiang}}, \bibinfo {author} {\bibfnamefont {Manuel}\
  \bibnamefont {Houzet}}, \bibinfo {author} {\bibfnamefont {Ram\'{o}n}\
  \bibnamefont {Aguardo}}, \bibinfo {author} {\bibfnamefont {Charles~M.}\
  \bibnamefont {Lieber}}, \ and\ \bibinfo {author} {\bibfnamefont {Silvano}\
  \bibnamefont {De~Franceschi}},\ }\bibfield  {title} {\enquote {\bibinfo
  {title} {{Spin-resolved Andreev levels and parity crossings in hybrid
  superconductor-semiconductor nano structures}},}\ }\href {\doibase
  10.1038/nnao.2013.267} {\bibfield  {journal} {\bibinfo  {journal} {Nature
  Nanotechnology}\ }\textbf {\bibinfo {volume} {9}},\ \bibinfo {pages} {79}
  (\bibinfo {year} {2014})}\BibitemShut {NoStop}%
\end{thebibliography}%

\end{document}